\documentclass[twoside,twocolumn,9pt]{article}
\usepackage{extsizes}
\usepackage[super,sort&compress,comma]{natbib} 
\usepackage[version=3]{mhchem}
\usepackage[left=1.5cm, right=1.5cm, top=1.785cm, bottom=2.0cm]{geometry}
\usepackage{balance}
\usepackage{mathptmx}
\usepackage{sectsty}
\usepackage{graphicx} 
\usepackage{lastpage}
\usepackage[format=plain,justification=justified,singlelinecheck=false,font={stretch=1.125,small,sf},labelfont=bf,labelsep=space]{caption}
\usepackage{float}
\usepackage{fancyhdr}
\usepackage{fnpos}
\usepackage[english]{babel}
\addto{\captionsenglish}{%
  
}
\usepackage{array}
\usepackage{droidsans}
\usepackage{charter}
\usepackage[T1]{fontenc}
\usepackage[usenames,dvipsnames]{xcolor}
\usepackage{setspace}
\usepackage[compact]{titlesec}
\usepackage{hyperref}
\usepackage{hyphenat}
\usepackage{dblfloatfix}

\usepackage{epstopdf}

\definecolor{cream}{RGB}{222,217,201}

\begin{document}

\pagestyle{fancy}
\thispagestyle{plain}
\fancypagestyle{plain}{
\renewcommand{\headrulewidth}{0pt}
}

\makeFNbottom
\makeatletter
\renewcommand\LARGE{\@setfontsize\LARGE{15pt}{17}}
\renewcommand\Large{\@setfontsize\Large{12pt}{14}}
\renewcommand\large{\@setfontsize\large{10pt}{12}}
\renewcommand\footnotesize{\@setfontsize\footnotesize{7pt}{10}}
\makeatother

\renewcommand{\thefootnote}{\fnsymbol{footnote}}
\renewcommand\footnoterule{\vspace*{1pt}%
\color{cream}\hrule width 3.5in height 0.4pt \color{black}\vspace*{5pt}} 
\setcounter{secnumdepth}{5}

\makeatletter 
\renewcommand\@biblabel[1]{#1}            
\renewcommand\@makefntext[1]%
{\noindent\makebox[0pt][r]{\@thefnmark\,}#1}
\makeatother 
\renewcommand{\figurename}{\small{Fig.}~}
\sectionfont{\sffamily\Large}
\subsectionfont{\normalsize}
\subsubsectionfont{\bf}
\setstretch{1.125} 
\setlength{\skip\footins}{0.8cm}
\setlength{\footnotesep}{0.25cm}
\setlength{\jot}{10pt}
\titlespacing*{\section}{0pt}{4pt}{4pt}
\titlespacing*{\subsection}{0pt}{15pt}{1pt}
\raggedbottom

\fancyfoot{}
\fancyfoot[LO,RE]{\vspace{-7.1pt}\includegraphics[height=9pt]{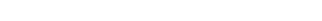}}
\fancyfoot[CO]{\vspace{-7.1pt}\hspace{13.2cm}\includegraphics{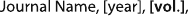}}
\fancyfoot[CE]{\vspace{-7.2pt}\hspace{-14.2cm}\includegraphics{head_foot/RF}}
\fancyfoot[RO]{\footnotesize{\sffamily{1--\pageref{LastPage} ~\textbar  \hspace{2pt}\thepage}}}
\fancyfoot[LE]{\footnotesize{\sffamily{\thepage~\textbar\hspace{3.45cm} 1--\pageref{LastPage}}}}
\fancyhead{}
\renewcommand{\headrulewidth}{0pt} 
\renewcommand{\footrulewidth}{0pt}
\setlength{\arrayrulewidth}{1pt}
\setlength{\columnsep}{6.5mm}
\setlength\bibsep{1pt}

\makeatletter 
\newlength{\figrulesep} 
\setlength{\figrulesep}{0.5\textfloatsep} 

\newcommand{\topfigrule}{\vspace*{-1pt}%
\noindent{\color{cream}\rule[-\figrulesep]{\columnwidth}{1.5pt}} }

\newcommand{\botfigrule}{\vspace*{-2pt}%
\noindent{\color{cream}\rule[\figrulesep]{\columnwidth}{1.5pt}} }

\newcommand{\dblfigrule}{\vspace*{-1pt}%
\noindent{\color{cream}\rule[-\figrulesep]{\textwidth}{1.5pt}} }

\makeatother

\twocolumn[
  \begin{@twocolumnfalse}
{\includegraphics[height=30pt]{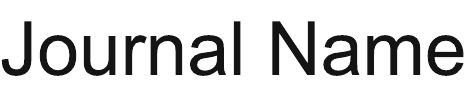}\hfill\raisebox{0pt}[0pt][0pt]{\includegraphics[height=55pt]{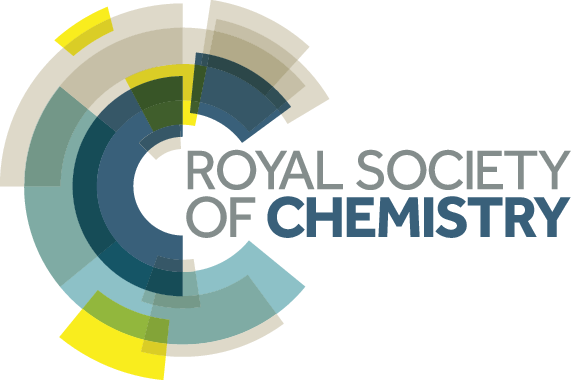}}\\[1ex]
\includegraphics[width=18.5cm]{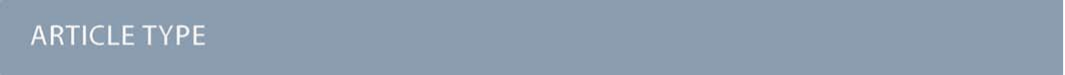}}\par
\vspace{1em}
\sffamily
\begin{tabular}{m{4.5cm} p{13.5cm} }

\includegraphics{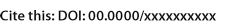} & \noindent\LARGE{\nohyphens{\textbf{Crystalline Morphology Formation in Phase-Field Simulations of Binary Mixtures$^\dag$}}} \\
\vspace{0.3cm} & \vspace{0.3cm} \\

 & \noindent\large{Maxime Siber,$^{\ast}$\textit{$^{a,b}$} Olivier J. J. Ronsin,\textit{$^{a}$} and Jens Harting\textit{$^{a,b,c}$}} \\

\includegraphics{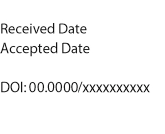} & \noindent\normalsize{Understanding the morphology formation process of solution-cast photoactive layers (PALs) is crucial to derive design rules for optimized and reliable third generation solar cell fabrication. For this purpose, a Phase-Field (PF) computational framework dedicated to the simulation of PAL processing has recently been developed. In this study focused on non-evaporating, crystallizing binary mixtures, distinct crystalline morphology formation pathways are characterized by a systematic exploration of the model's parameter space. It is identified how, depending on material properties, regular, dilution-enhanced, diffusion-limited and demixing-assisted crystallization can take place, and which associated structures then arise. A comprehensive description of the thermodynamic and kinetic mechanisms that respectively drive these separate crystallization modes is provided. Finally, comparisons with experimental results reported in the literature highlight the promising potential of PF simulations to support the determination of process-structure relationships for improved PAL production.} \\

\end{tabular}

 \end{@twocolumnfalse} \vspace{0.6cm}

  ]

\renewcommand*\rmdefault{bch}\normalfont\upshape
\rmfamily
\section*{}
\vspace{-1cm}


\footnotetext{\textit{$^{a}$~Helmholtz Institute Erlangen-Nürnberg for Renewable Energy, Forschungszentrum Jülich, Fürther Straße 248, 90429 Nürnberg, Germany, E-mail: m.siber@fz-juelich.de}}
\footnotetext{\textit{$^{b}$~Department of Chemical and Biological Engineering, Friedrich-Alexander-Universität Erlangen-Nürnberg, Fürther Straße 248, 90429 Nürnberg, Germany}}
\footnotetext{\textit{$^{c}$~Department of Physics, Friedrich-Alexander-Universität Erlangen-Nürnberg, Fürther Straße 248, 90429 Nürnberg, Germany}}

\footnotetext{\dag~Electronic Supplementary Information (ESI) available: [details of any supplementary information available should be included here]. See DOI: 00.0000/00000000.}




\section{Introduction}\label{Sec:Intro}

With persistently increasing power conversion efficiencies (PCEs) and a broadening field of applications, solution-processed third-generation solar cells are expected to play a significant part in the transition towards more sustainable and environment-friendly energy production infrastructures. Especially photovoltaic devices manufactured from perovskite and/or organic semiconductors present promising perspectives for both low-cost and energy-efficient fabrication on an industrial scale \cite{li_scalable_2018} \cite{brabec_material_2020}. Indeed, already well established solvent-based thin film deposition techniques, such as slot-die coating, are compatible with these materials and can be integrated in a roll-to-roll process \cite{yang_upscaling_2021} \cite{sondergaard_roll--roll_2013}. Nonetheless, despite attractive processing features, both technologies still face numerous challenges on the path towards reliable lab-to-factory scale-up and widespread commercialization \cite{williams_current_2016} \cite{wang_recent_2021}. Although the nature of the confronted issues is specific to the selected material properties, the nanostructure of the photoactive layer (PAL) is always determinant for the entire device characteristics, and, most importantly, for PCE. Since the morphology of the PAL arises during film fabrication, understanding how it is controlled by process parameters is of paramount importance to produce performant and robust solar cells.

At present, advances regarding optimal PAL processing are still mainly achieved by following a time- and resource-costly trial-and-error approach and many process-structure relationships remain yet to be unraveled. Specialized numerical simulations are therefore of interest to gain theoretical insights into the complex interaction of physical phenomena that shapes the PAL at nanometric scale \cite{jackson_coarse-graining_2021}. For instance, organic photovoltaic (OPV) material systems are likely to experience a miscibility gap while undergoing temperature and composition changes during processing \cite{mcdowell_solvent_2018} \cite{peng_materials_2023}. Knowing under which conditions amorphous-amorphous phase separation (AAPS) interferes with other structure formation mechanisms is thereby of crucial interest to pinpoint the best way for device fabrication. For this purpose, multiphysics Phase-Field (PF) modelling frameworks were employed by Wodo et al. \cite{wodo_modeling_2012}, Kouijzer et al. \cite{kouijzer_predicting_2013}, and Schaefer et al. \cite{schaefer_structuring_2016}, to examine the coupling of AAPS and evaporation in ternary solvent-donor-acceptor blends for OPV. Furthermore, single component crystallization in OPV films produced via meniscus-guided coating was also addressed by Michels and co-workers \cite{michels_predictive_2021}. Finally, a PF framework combining all these phenomenologies was recently designed to compute the time-resolved evolution of solution-cast PAL architectures over the whole fabrication process \cite{ronsin_phase-field_2022}. In previous reports, most relevant mechanisms involved in the manufacturing of bulkheterojunction-type PALs (BHJ), namely spinodal decomposition (SD) \cite{konig_two-dimensional_2021}, its interplay with crystal growth \cite{ronsin_role_2020}, and solvent evaporation \cite{ronsin_phase-field_2020} \cite{ronsin_phase-field_2021} were successfully implemented and analysed in-depth. A complete investigation of the film drying, which encompasses all aforementioned physics, was also carried out for the extensively surveyed P3HT-PCBM OPV model system \cite{ronsin_formation_2022}.

A comprehensive study of the progress of crystallization along available morphology formation pathways, which are recognized to be strongly impacted by competing processes such as AAPS \cite{wang_coupling_2021}, remains to be realized. Therefore, it is now sought to obtain a general overview of the crystallization based on the physics that are captured by this PF framework. More precisely, the objective is to understand how material and processing parameters influence the crystallization kinetics and the resulting morphologies. The scope is restricted to non-evaporating incompressible binary mixtures with only one crystallizing species, as this configuration readily provides numerous complex cases for systematical analysis. Nevertheless, the application range extends beyond the material classes used in the field of solution-processed photovoltaics. In this broader context, the PF method is a well-established continuum mechanics simulation approach that allows separately for modelling of crystallization \cite{granasy_phase-field_2019} and amorphous-amorphous phase separation (AAPS) \cite{wodo_computationally_2011}. Different couplings between both phenomena have been derived to investigate crystallization processes in specific systems such as nucleation assisted by phase separation fluctuation in PEH-PEB polymer-polymer blends \cite{zhou_numerical_2008}, metal alloy solidification from regular and phase separating solutions \cite{granasy_phase-field_2019} or microstructure evolution in controlled drug-release coatings \cite{kim_modeling_2009} \cite{saylor_predicting_2011} \cite{saylor_diffuse_2016}. A development, which the current code relies on, has been proposed by Matkar and Kyu \cite{matkar_role_2006} to account simultaneously for crystallization and AAPS and permit supplementary crystalline-amorphous and crystalline-crystalline demixing. In this work, two main tasks are associated with the overall aim: First, a parametric study is conducted to elucidate the sensitivity of the crystallization dynamics against thermodynamic and kinetic parameter variations and reveal possible pathways for crystalline morphology formation. Second, the identified distinct crystallization modes are further categorized with respect to the underlying physical properties that activate them.

The remainder of this paper is organized as follows: Sec.\ref{Sec:PFModel} focuses on the description of the PF model while Sec.\ref{Sec:Theory} supports the theoretical analysis of the simulated crystallization behavior with discussions about phase equilibria and potential nucleation pathways through the free energy landscape. Sec.\ref{Sec:CrystallizationBehavior} then reports both the morphology and the crystallization kinetics observed for the initial OPV reference simulation setup in the regular crystallization mode. Additionally, results from the sensitivity study are presented here. In Sec.\ref{Sec:CrystallizationModes}, diffusion-limited, demixing-assisted and dilution-enhanced crystalline structure formation scenarios are successively characterized and noteworthy arising morphological features are detailed. A short literature survey focusing on OPV systems is undertaken in Sec.\ref{Sec:Discussion} to validate simulation outcomes against already published experimental results. Finally, the conclusions of this work are presented in Sec.\ref{Sec:Conclusion}.

\section{Description of the Phase-Field Model}\label{Sec:PFModel}

\subsection{The Free Energy Functional}

The PF approach makes use of field variables to represent the local phase properties of the investigated material system. Since the PF variables vary continuously across the simulation domain, interfaces between different phases that arise within the system are inherently diffuse. Here, with the aim to simulate crystallization in incompressible binary blends where only one component can crystallize, two different parameters denoted by $\phi$ and $\psi$ are employed. On the one hand, $\phi$ is a compositional PF variable which monitors the volume fraction of the crystalline component. Even though the scope of this study is broader than solvent-solute mixtures, the crystalline species is abusively referred to as "solute" in what follows for simplicity. In contrast, the amorphous one is labeled as "solvent". On the other hand, $\psi$ is a structural order parameter which tracks the extent of the crystallization transformation. It also ranges from 0 (fully amorphous state) to 1 (fully ordered state, i.e completed crystallization). 

Depending on the values of the field variables $\phi$ and $\psi$, successive evaluations of the system's free energy dictate how it evolves in time. The PF model used in this article is a particular case of a more general previously established framework \cite{ronsin_phase-field_2022} to focus on the coupled physical phenomena that are relevant for the present problem, namely crystallization and phase separation. The former can be defined as the transition from a micro-structurally disordered phase, termed amorphous phase, to an energetically more favorable phase with compact ordered micro-structure: the crystal. For a crystal to form, an initial localized energy fluctuation is always required to enable the system to build a stable nucleus that overcomes the crystallization energy barrier and start free crystal growth \cite{granasy_phase-field_2019}. For this reason, crystallization is categorized as a nucleation and growth (NG) process. At late crystallization stages, free growth can be hindered by crystal impingement, lack of material, inadequate diffusion properties or other boundary conditions. 

The second process, phase separation, is the demixing of a blend of at least two species which mixed state is thermodynamically unstable (or metastable). As a result, distinct, well-defined domains, which are respectively richer in each of the separating components, are formed within the system. Differently from crystallization, phase separation can either be triggered locally through NG when the mixture is metastable, or happen spontaneously and simultaneously in the whole blend when it is unstable. In the latter case, the demixing is referred to as spinodal decomposition (SD). In addition, it can be mentioned that species involved in the dissociation process might be in the crystalline or in the amorphous state. In this report, amorphous-amorphous phase separation is abbreviated with AAPS.

In order to describe the thermodynamics of both phase change phenomena, the Gibbs free energy $G$ of the system is written as the sum of two corresponding contributions,

\begin{equation}
    G = G^{(ac)} + G^{(mix)} ~,
    \label{eq:EnergySum}
\end{equation}

where $G^{(ac)}$ represents the free energy change due to transitions from the amorphous to the crystalline state, while $G^{(mix)}$ captures interactions between both mixed components that determine whether the system is prone to phase separation or not. Following the treatment by Cahn and Hilliard \cite{cahn_free_1958} or Ginzburg and Landau \cite{hohenberg_introduction_2015}, the free energy of the system is typically expressed as a functional in the context of PF modelling. The next subsections detail further how both terms of Eq.\ref{eq:EnergySum} are formulated in the employed framework.

\subsubsection{Free Energy of Crystallization}

The free energy change due to crystallization ($G^{(ac)}$ in Eq.\ref{eq:EnergySum}) is rewritten as a functional that includes three main contributions:

\begin{equation}
    G^{(ac)} = \int_V \left[ G_V^{(bulk,ac)}(\phi,\psi) + G_V^{(grad,ac)}(\nabla \psi) + G_V^{(ori,ac)}(\theta, \nabla \theta) \right] dV ~,
    \label{eq:CrystallizationFunctional}
\end{equation}


whereby, for the present PF simulations, $V$ indicates the volume of the discrete mesh elements which the simulation box consists of. The first term in Eq.\ref{eq:CrystallizationFunctional} accounts for the bulk free energy of crystallization phase change and has to take the form of a double-well potential in order to mimic the behavior of NG processes. Usually, polynomial interpolation functions are exploited to provide this property \cite{granasy_phase-field_2019}. Here, a combination of a quartic function $q(\psi)= \psi^2(1-\psi)^2$, which creates the desired double-well shape, and a cubic interpolation $p(\psi) = \psi^2(3-2\psi)$, that regulates the latent heat release with increasing order parameter $\psi$, is employed. The polynomial $q(\psi)$ is proportional to a coefficient $W$ (in units J/kg) which is related to the crystal's surface tension and controls the height of the energy barrier to be overcome before NG. Likewise, $p(\psi)$ is weighted by a factor $L(T/T_m - 1)$, which stems from Turnbull's linear approximation of the free energy decrease for isotropically crystallizing species at low undercoolings \cite{granasy_phase-field_2019}. Analogously for lamellar polymer crystallization, a similar factor emerges in Lauritzen and Hoffmann's treatment \cite{lauritzen_theory_1960}. $L$ (in J/kg as well) represents the latent heat of fusion (assumed here to be constant), while $T$ and $T_m$ respectively denote the system's temperature and the melting temperature of the crystalline component. The total volume specific bulk free energy change due to crystallization finally reads

\begin{equation}
    G_V^{(bulk,ac)} = \phi \rho \left[q(\psi)W + p(\psi) L \left( \frac{T}{T_m}-1 \right) \right] ~,
    \label{eq:CrystallizationHomogeneous}
\end{equation}

with $\rho$ the density of the species subject to crystallization. 

Furthermore, the second term in Eq.\ref{eq:CrystallizationFunctional} describes the energy rise pertaining to transitions in molecular ordering at crystal boundaries:

\begin{equation}
    G_V^{(grad,ac)} = \frac{\varepsilon^2}{2} |\nabla\psi|^2~.
    \label{eq:CrystallizationGradients}
\end{equation}

Therefore, it scales with the squared gradient of the order parameter $|\nabla \psi|^2$ as well as a coefficient $\varepsilon^2$, which, similarly to $W$, is linked to the crystal surface tension \cite{takaki_phase-field_2014}. 

The last term in Eq.\ref{eq:CrystallizationFunctional} introduces a supplementary PF variable $\theta$ that attributes an orientation angle to nucleated crystals and writes 

\begin{equation}
    G_V^{(ori,ac)} = p(\psi) \frac{\alpha}{2} \delta_{\nabla \theta}~.
    \label{eq:CrystallizationOrientation}
\end{equation}

The parameter $\theta$ is required to handle crystal impingement due to orientation mismatch and prevent interpenetration. An energy coefficient $\alpha$ scales the related energy contribution. The Kronecker symbol $\delta_{\nabla \theta}$ indicates that the energy rise is only active if a gradient in $\theta$ exists in the crystal phase. The interpolation polynomial $p(\psi)$ then regulates the strength of $G_V^{(ori,ac)}$ according to the advancement of crystallization at the impingement location. Note that $\theta$ can be used to implement anistropic crystal growth \cite{granasy_phase-field_2014}, which is often observed for polymers and organic small molecules. Here, however, the crystals were maintained isotropic since anisotropy is not expected to impact the trends in crystallization kinetics significantly.

\subsubsection{Free Energy of Mixing}

Analogously to the crystallization term, the free energy change due to mixing properties can be split into several contributions:  

\begin{equation}
    G^{(mix)} = \int_V \left[ G_V^{(bulk,mix)}(\phi,\psi) + G_V^{(grad,mix)}(\nabla \phi) + G_V^{(num,mix)}(\phi) \right] dV ~.
    \label{eq:MixingFunctional}
\end{equation}

Again, $G_V^{(bulk,mix)}$ and $G_V^{(grad,mix)}$ designate bulk and interfacial free energy densities, respectively, whereas $G_V^{(mix,num)}$ is a supplemented term designed to ensure numerical stability when volume fractions reach values close to 0 or 1 \cite{kim_modeling_2009} \cite{ronsin_phase-field_2022}. This contribution does not affect significantly the physics of the simulations and is therefore not detailed further in this article.

According to the Flory-Huggins theory of mixing \cite{flory_principles_1953}, which allows modelling of small molecule solutions up to polymer mixtures, and its generalization by Matkar and Kyu \cite{matkar_role_2006} to include crystalline components, the bulk contribution $G_V^{(bulk,mix)}$ can further be written as

\begin{equation}
\begin{split}
    G_V^{(bulk,mix)} = \frac{RT}{v_0} & \left[ \frac{ \phi \ln{(\phi)}}{N_1} + \frac{(1-\phi) \ln{(1-\phi)}}{N_2} \right. \\
    & \left. + \phi (1-\phi) (\chi_{aa} + \chi_{ca}\psi^2) \vphantom{\frac{ \phi \ln{(\phi)}}{N_1}} \right] ~.
    \label{eq:MixingBulk}
\end{split}
\end{equation}

Here, $R$ is the ideal gas constant and $v_0$ the molar volume of the Flory-Huggins lattice elements, which can be as large as atoms, monomers, polymer segments or small molecules, depending on the selected material system. $N_1$ and $N_2$ then express the respective species sizes in terms of these elements. Throughout this report, subscript 1 refers to the crystallizing component and 2 to the amorphous one. Since the amorphous species always possesses the smallest molar volume in the present investigations, $v_0$ is indexed on it and $N_2$ is equal to 1.

$\chi_{aa}$ denotes the so-called Flory-Huggins interaction parameter which accounts for pairwise interactions between neighbouring Flory-Huggins lattice elements in the amorphous state. The nature of these interactions, which determine the miscibility of the amorphous mixture, may be enthalpic or entropic. In general, a linear relationship in $1/T$ is assumed for the temperature-dependence of $\chi_{aa}$ \cite{rubinstein_polymer_2003}:

\begin{equation}
    \chi_{aa} = A + \frac{B}{T} ~.
    \label{eq:AAInteraction}
\end{equation}

The offset $A$ is associated with entropic contributions while the slope $B$ sets the enthalpic interaction trend in $1/T$. In experiments, different methods are commonly employed to evaluate $\chi_{aa}$ \cite{kouijzer_predicting_2013} \cite{perea_introducing_2017} \cite{ye_quantitative_2018} \cite{yao_preaggregation_2022}. Nevertheless, adjustment of the classical Flory-Huggins model was suggested by Matkar and Kyu \cite{matkar_role_2006} to account for separate interaction contributions that take place between components in the amorphous and the ordered state. To this end, an auxiliary interaction parameter $\chi_{ca}$ that satisfies the following relation was introduced:

\begin{equation}
    \chi_{ca} = C \frac{v_0 N_1 \rho L}{RT} ~.
    \label{eq:CAInteraction}
\end{equation}

In this expression, the constant $C$ is a proportionality factor. It can be mentioned that this is in qualitative agreement with the Flory-Huggins theory modification by Hu \cite{hu_polymer_2013} that also results in a corrected Flory-Huggins interaction parameter with a supplementary term proportional to a crystallization energy. Additionally, the multiplication of $\chi_{ca}$ with $\psi^2$ in Eq.\ref{eq:MixingBulk} couples ordering transitions to composition changes, so that crystallization of one species is likely to be accompanied by a demixing process. Besides temperature, $\chi_{aa}$ and $\chi_{ca}$ are also expected to change with blend composition \cite{flory_principles_1953}. For simplicity, they are however assumed independent of $\phi$ in this work.

Finally, like the order parameter $\psi$, gradients in volume fraction $\phi$ at phase interfaces give rise to an energy contribution related to surface tension. The corresponding term, $G_V^{(grad,mix)}$ in Eq.\ref{eq:MixingFunctional}, scales with a coefficient $\kappa$ which is comparable to $\varepsilon^2$ in $G_V^{(grad,ac)}$ (Eq.\ref{eq:CrystallizationGradients}):

\begin{equation}
    G_V^{(grad,mix)} = \frac{\kappa}{2}|\nabla \phi|^2 ~.
    \label{eq:MixingGradients}
\end{equation}

\subsection{Kinetic Transport Equations}

\subsubsection{Stochastic Allen-Cahn Equation}

In order to track the time-evolution of non-conserved order parameter fields, the Allen-Cahn equation is usually employed \cite{allen_microscopic_1979}. For the purpose of this work it is augmented with a Langevin force term to reproduce the effect of thermal fluctuations on the system. Under this form, it is known as stochastic Allen-Cahn equation and reads as

\begin{equation}
    \frac{\partial \psi}{\partial t} = - \frac{v_0 N_1}{RT} M(\phi) \frac{\delta G}{\delta \psi} + \xi ~.
    \label{eq:AC}
\end{equation}

Here, $\delta G/\delta \psi$ symbolizes the functional derivative of the free energy with respect to $\psi$, which, once written in terms of the corresponding partial derivatives of the volume specific free energy, yields

\begin{equation}
    \frac{\partial \psi}{\partial t} = -\frac{v_0 N_1}{RT} M(\phi) \left[ \frac{\partial G_V}{\partial \psi} -\nabla \cdot \frac{\partial G_V}{\partial (\nabla \psi)} \right] + \xi ~.
    \label{eq:ACextended}
\end{equation}

$M(\phi)$ denotes a kinetic parameter that determines the local conversion rate of amorphous material into crystalline one. Therefore, it is determinant for crystal nucleation properties and interface mobilities during growth. As it is further detailed in an upcoming section (Sec.\ref{Sec:DilEn}), $M$ is expected to be composition-dependent \cite{kim_modeling_2009} in the general case. Finally, the Langevin term $\xi$ is a Gaussian noise with a standard deviation of $\frac{2v_0}{N_a}N_1 M(\phi)$ designed to preserve the fluctuation-dissipation theorem \cite{shen_effect_2007}. It is responsible for thermally activated nucleation, as well as crystal coarsening.

\subsubsection{Cahn-Hilliard Conservation Equation}

Conserved quantities such as the volume fraction $\phi$ satisfy a Cahn-Hilliard-type continuity equation \cite{lee_physical_2014} which equates their partial derivative with respect to time to the divergence of their corresponding flux $\mathbf{J}$. Considering assumptions from non-equilibrium thermodynamics theory for multicomponent systems \cite{groot_non-equilibrium_1984}, $\mathbf{J}$ can be substituted as follows:  

\begin{equation}
    \frac{\partial \phi}{\partial t} = - \nabla \cdot \mathbf{J} = - \frac{v_0}{RT}  \nabla \cdot \left[ \Lambda(\phi) \nabla (\mu_1-\mu_2) \right] ~.
    \label{eq:CH}
\end{equation}

The quantity $\mu_1-\mu_2$ is a chemical potential density which is sometimes referred to as exchange chemical potential \cite{kouijzer_predicting_2013} \cite{peng_materials_2023} and which derives from the classical diffusion potential \cite{hillert_phase_2008} \cite{ronsin_phase-field_2021}. Its gradient is the thermodynamic driving force for mass diffusion. According to Cahn \cite{cahn_phase_1965}, it can be computed as the functional derivative of the free energy with respect to $\phi$, i.e.

\begin{equation}
    \mu_1-\mu_2 = \frac{\delta G}{\delta \phi} = \frac{\partial G_V}{\partial \phi} - \nabla \cdot \frac{\partial G_V}{\partial(\nabla \phi)} ~.
    \label{eq:DiffusionPotential}
\end{equation}

Furthermore, $\Lambda(\phi)$ designates the Onsager mobility coefficient \cite{onsager_reciprocal_1931}, which, in the more general multicomponent case, takes the form of a positive semi-definite matrix. Diffusion of species within the system is regulated by this kinetic parameter. To calculate $\Lambda$ depending on $\phi$, the PF code implements the fast \cite{kramer_interdiffusion_1984} and slow \cite{de_gennes_dynamics_1980} mode theories. Both have been confirmed to lack generality so that attempts to unify them have been published \cite{akcasu_fast_1997}. However, the fast mode is selected here, as a strong impact of the chosen theory on the qualitative trends described in this article is not expected. $\Lambda(\phi)$ accordingly writes as

\begin{equation}
    \Lambda(\phi) = (1-\phi)^2 \phi N_1 D_1^{\mathrm{(self)}}(\phi) + \phi^2 (1-\phi) N_2 D_2^{\mathrm{(self)}}(\phi) ~,
    \label{eq:Lambda}
\end{equation}

where $D_1^{\mathrm{(self)}}(\phi)$ and $D_2^{\mathrm{(self)}}(\phi)$ stand for the self-diffusion coefficients of species 1 and 2, respectively. To interpolate their values in the volume fraction range from measurements in pure materials ($D_{1|{\phi\rightarrow 0}}^{\mathrm{(self)}}$, $D_{2|{\phi\rightarrow 0}}^{\mathrm{(self)}}$, $D_{1|{\phi\rightarrow 1}}^{\mathrm{(self)}}$, $D_{2|{\phi\rightarrow 1}}^{\mathrm{(self)}}$), a logarithmic mean is utilized for simplicity (see supporting information SI-A). 

\begin{equation}
    \begin{cases}
        D_1^{\mathrm{(self)}}(\phi) = & \left( D_{1|{\phi\rightarrow 0}}^{\mathrm{(self)}} \right)^{(1-\phi)} \left( D_{1|{\phi\rightarrow 1}}^{\mathrm{(self)}} \right)^{\phi}~,\\
        D_2^{\mathrm{(self)}}(\phi) = & \left( D_{2|{\phi\rightarrow 0}}^{\mathrm{(self)}} \right)^{\phi} \left( D_{2|{\phi\rightarrow 1}}^{\mathrm{(self)}} \right)^{(1-\phi)} ~.
    \end{cases}
    \label{eq:SelfD}
\end{equation}

Note that other dependencies have been suggested, as estimating the composition-dependence of self-diffusion coefficients is still an open research topic \cite{wolff_prediction_2018}.

Similarly to the stochastic Allen-Cahn transport equation, fluctuations can be added in the Cahn-Hilliard formula (Eq.\ref{eq:CH}), which is then referred to as Cahn-Hilliard-Cook equation \cite{cook_brownian_1970}. These fluctuations are mainly of use to trigger AAPS. It was noticed during preliminary experiments that fluctuations in the stochastic Allen-Cahn equation (Eq.\ref{eq:AC}) already serve this purpose due to the coupling between crystallization and phase separation free energies (Eq.\ref{eq:MixingBulk}). The noise contribution on the Cahn-Hilliard equation is therefore omitted to ensure better numeric robustness, especially regarding volume fraction conservation.

\section{Theoretical Analysis}\label{Sec:Theory}

\subsection{Phase Equilibria}\label{Sec:PhaseDiagrams}

In accordance with Gibbs' phase rule, binary systems under constant pressure and temperature conditions can present at most two distinct phases in the equilibrium state. Since the PF model relies on a free energy formulation, information about the phase equilibrium compositions can already be obtained before carrying out numerical simulations of the system's evolution in time. Here, the approach described by Horst \cite{horst_calculation_1995} \cite{horst_calculation_1996} is employed in order to produce $\phi-T$ phase diagrams. Note that, although solely this sort of graphs is featured in this article, this routine can as well be used to trace figures with different ordinate axes, as for instance $\phi-\chi_{aa}$ diagrams, which are also often examined in the literature \cite{kozub_polymer_2011} \cite{ye_quantitative_2018} \cite{zhu_rational_2019} \cite{konig_two-dimensional_2021} \cite{peng_materials_2023}.

\begin{figure}[!htb]
    \centering
    \includegraphics[scale=0.4]{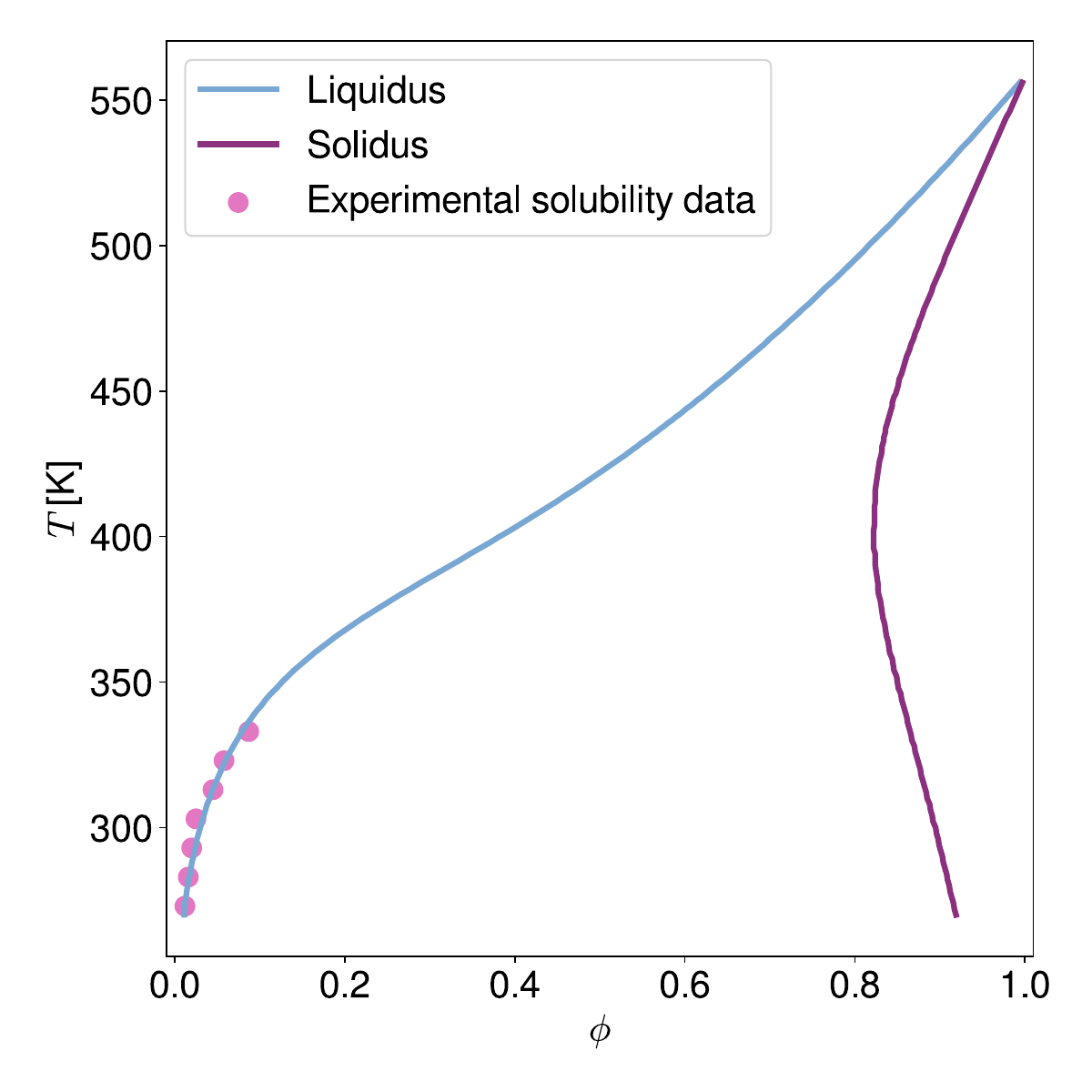}
    \caption{Phase diagram for the PCBM-oDCB system as extrapolated with the current free energy model from experimental solubility measurements by Schmidt-Hansberg et al. \cite{schmidt-hansberg_moving_2011} ($\chi_{aa}$ = 0.34 + 128.14/$T$ and $\chi_{ca}$~=~0.1648$\frac{v_0 N_1 \rho L}{RT}$). Other thermodynamic parameters used for the computation are listed in Tab.\ref{tab:SimulationParameter}.}
    \label{fig:PhaseDiagram}
\end{figure}

The typical form of the phase diagram corresponding to systems investigated in the present work can be visualized in Fig.\ref{fig:PhaseDiagram}. The increasing divergence of the liquidus and the solidus from the pure solute melting temperature towards lower $T$ is notably due to the UCST-type (upper critical solution temperature) behavior induced by $\chi_{aa}$ (with positive coefficient $B$ in Eq.\ref{eq:AAInteraction}). Even though multitudes of diagram shapes, including LCST-types (lower critical solution temperature), can be achieved with this free energy \cite{matkar_phase_2006}, this sort of diagram is considered as a well-suited starting point for the parametric study since it is frequently reported in the OPV literature \cite{ye_miscibilityfunction_2018} \cite{ghasemi_delineation_2019} \cite{wadsworth_bulk_2020} \cite{peng_materials_2023}.

The reference parameter set that corresponds to the phase diagram depicted in Fig.\ref{fig:PhaseDiagram} represents a mixture of (6,6)-phenyl C61 butyric acid methyl ester (PCBM) dissolved in ortho-dichlorobenzene (oDCB). This OPV small molecule-solvent blend is selected as a practical initial system due to the availability of experimental data from which inputs for the PF model can be derived \cite{ronsin_formation_2022}. In the OPV community, PCBM is well-known as the electron acceptor of the extensively studied P3HT-PCBM polymer-small molecule bulkheterojunction (BHJ) \cite{ludwigs_p3ht_2014} \cite{berger_polymer_2018} \cite{wadsworth_bulk_2020}. More recently, PCBM has also been mixed with non-fullerene acceptors (NFAs) in ternary and quaternary OSCs to improve device performance \cite{guo_recent_2021}. Ortho-dichlorobenzene (oDCB) is a high-boiling point solvent commonly used for its affinity with a wide variety of OPV materials \cite{mcdowell_solvent_2018}.

Thermodynamic parameters employed for the calculation of PCBM-oDCB phase equilibria are summarized in Tab.\ref{tab:SimulationParameter} and have been discussed in detail in previous numerical investigations of this system \cite{ronsin_formation_2022}. In this work, the energy barrier coefficient $W$ is slightly modified to grant the necessary double-well shape of the crystallization free energy over a sufficiently large parameter range for the sensitivity study (see limiting criterion on $W$ in SI-C). Moreover the interaction parameters $\chi_{aa}$ and $\chi_{ca}$ are adapted to fit solubility limit measurements published by Schmidt-Hansberg et al. \cite{schmidt-hansberg_moving_2011}. A qualitative summary of the impact of thermodynamic parameter variations on liquidus and solidus compositions is given in the supporting information (SI-B). It can be seen that an increase in both $\chi_{aa}$ and $\chi_{ca}$ shifts the position of the liquidus in opposite directions along the $\phi$-axis. Therefore, several [$\chi_{aa}$, $\chi_{ca}$] pairs could be used to extrapolate virtually valid diagrams from the experimental data points. The availability of the exact PCBM-oDCB phase diagram is not a mandatory requirement for the present endeavor. The initial parameter set is thus selected so that enough margin is provided for the parameter variations of the sensitivity study and to avoid numerical convergence issues \cite{takaki_phase-field_2014} \cite{ronsin_phase-field_2022}.  

\begin{figure*}[!htb]
    \centering
    \includegraphics[scale=0.4]{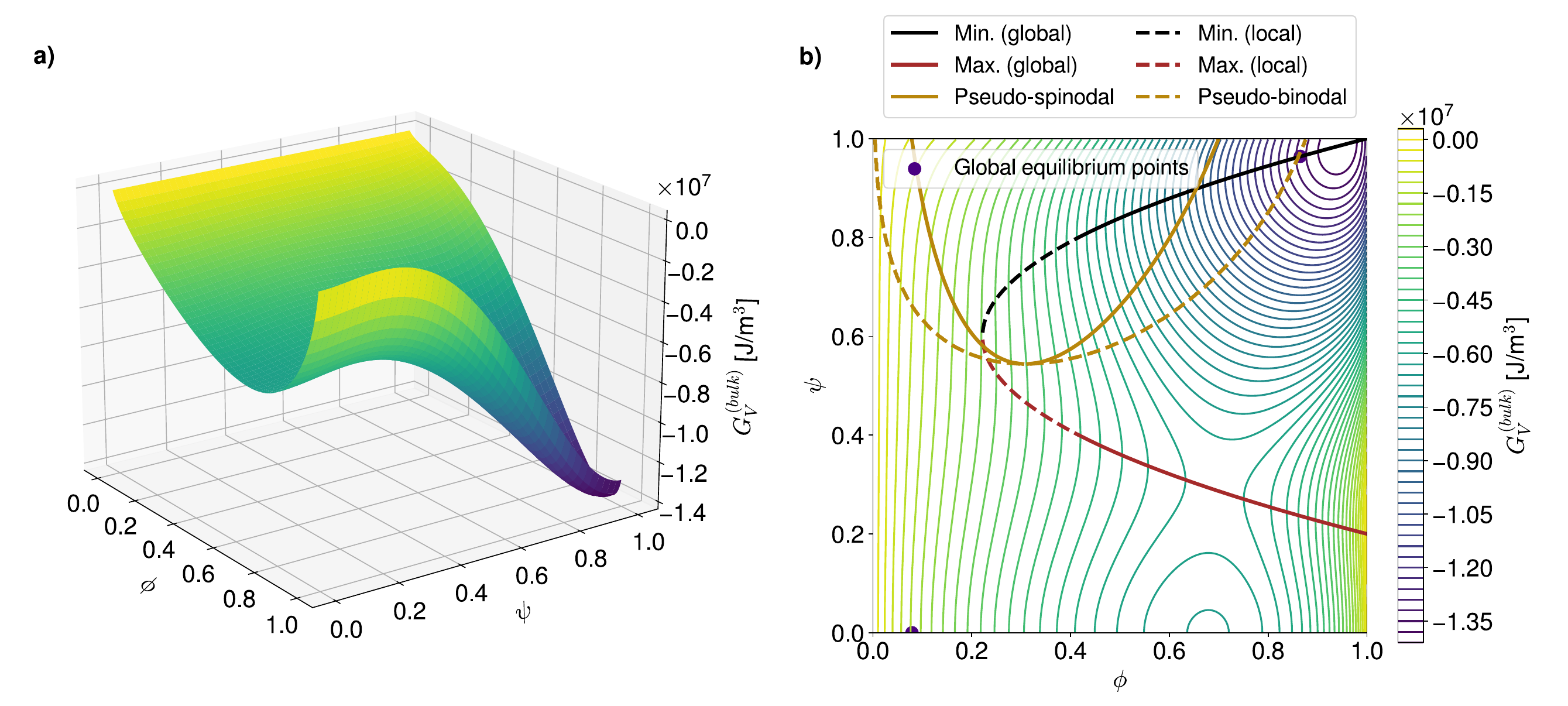}
    \caption{\textbf{a)} Three-dimensional free energy landscape computed for the PCBM-oDCB system at $T$ = 333 K. \textbf{b)} Corresponding 2D contour plot with highlighted demixing and crystal growth regions.}
    \label{fig:Landscape}
\end{figure*}

\subsection{Nucleation Pathways through the Free Energy Landscape}\label{Sec:Landscape}

The analysis of phase diagrams allows to predict the equilibrium states towards which thermodynamic systems eventually evolve. However, these diagrams contain no information about intermediate states which are explored on the way to the final equilibrium. This renders it difficult to ascertain under what conditions nuclei effectively appear and grow. In order to gain further understanding of the privileged routes for phase change at a given process temperature, the topology of the corresponding free energy surface can be examined. Fig.\ref{fig:Landscape}-a shows an example of such a three-dimensional free energy landscape computed for the PCBM-oDCB small molecule-solvent blend at 60\textdegree C. In the following, crystalline germ formation is considered from an originally amorphous blend (i.e. $\psi = 0$) with homogeneous composition $\phi_0$. The trigger for crystal nucleation is the stochastic term $\xi$ in the Allen-Cahn equation (Eq.\ref{eq:ACextended}), which is responsible for the emergence of regions with a non-zero order parameter $\psi$ in the mixture.

The first possible nucleation pathway through the landscape allows for stable crystallites to be produced in a single step. More precisely, if, over an area wider than the critical nucleus size, fluctuations on $\psi$ arise large enough to overcome the energy barrier, then crystal ordering (i.e. an increase in $\psi$), growth and purification in terms of solute content (i.e an increase in $\phi$) can spontaneously proceed. The second pathway is a two-step process which involves a demixing stage after the initial order parameter fluctuation and before the stable crystal build-up. To understand when each nucleation route is followed, the contour plot of the free energy landscape (Fig.\ref{fig:Landscape}-b) can be used. In Fig.\ref{fig:Landscape}-b, the red and black lines respectively stand for the maximum (i.e. the energy barrier) and minimum of the free energy at fixed solute volume fractions (see SI-C for the corresponding analytical formulae). The final, global equilibrium points of the thermodynamic system, which correspond to the liquidus and solidus in the phase diagram, are reproduced in purple. 

The one-step pathway is only available if the crystallization occurs in a mixture with a blend ratio $\phi_0$ for which an energy barrier exists (in the example of Fig.\ref{fig:Landscape}-b this corresponds to $\phi_0$ > 0.2). There, if, over a domain that is larger than the size of the critical nucleus, simultaneous local fluctuations surpass the energy barrier (see Sec.\ref{Sec:PFModel}), a crystal germ can form directly. Moreover, if, beyond the barrier, the minimum associated with the overall composition $\phi_0$ is a global minimum (solid black line in Fig.\ref{fig:Landscape}-b), the germ is stable and can only evolve towards the solidus point. This occurs by further crystal ordering and purification (i.e. increases in $\psi$ and $\phi$, respectively). In case the minimum after the barrier is local (dashed black line in Fig.\ref{fig:Landscape}-b), the resulting germ relies on composition changes due to crystal purification to stabilize. Therefore, if crystal purification is hindered for example due to strong diffusion limitations (see Sec.\ref{Sec:DiffLim}), one-step crystallization can be kinetically impeded below a threshold volume fraction (here $\phi_0 \leq$  0.4).

In comparison, two-step nucleation can take place when the initial fluctuation locally drives the system into a region of the landscape where the composition $\phi_0$ is energetically unstable or metastable (represented by the solid and dashed golden lines in Fig.\ref{fig:Landscape}-b, see SI-C for analytical derivation). The demixing is undergone in a phase dissociation process similar to AAPS with spinodal- and binodal-like behavior. It can be remarked that the regular AAPS in immiscible amorphous blends is in fact a particular case of this more general configuration where the strength of the fluctuations required to enter the demixing region is minimal. A free energy landscape for such a system can be visualized in Fig.\ref{fig:SpinodalPhaseDiagram}-b of Sec.\ref{Sec:AAPSCryst}. However, unlike common AAPS, the unstable (or metastable) state is in general not homogeneously reached in the whole system due to the local nature of the order parameter fluctuations. Thus, the demixing process is also less homogeneous.  While the pseudo-AAPS develops towards the compositions predicted by the binodal-like curves, crystallization can start as soon as the region delimited by the energy barrier and the minimum (red and black lines in Fig.\ref{fig:Landscape}-b) is entered. The exact volume fraction at which nucleation starts then depends on the balance between the kinetics related to the Cahn-Hilliard (Eq.\ref{eq:CH}) and Allen-Cahn (Eq.\ref{eq:AC}) equations.

For the particular PCBM-oDCB reference system, for which the landscape is shown in Fig.\ref{fig:Landscape}-a and Fig\ref{fig:Landscape}-b, it can be noted that the energy barrier is located at relatively low $\psi$ values and spans over the majority of the $\phi$-range ($\phi \in$ [0.2, 1.0]). In contrast, the AAPS-like region requires high fluctuations on $\psi$ to be attained. The two-step pathway is therefore much less probable than the one-step counterpart in this case. Below $\phi \simeq$ 0.2, where solely two-step processes can lead to crystallization, the appearance of nuclei is additionally strongly delayed. Indeed, at these lower overall solute concentrations, the solute majority phase which emerges from the demixing, and which reaches the landscape region that is beyond the energy barrier, is likely to be smaller than the critical nucleus size, so that, in the end, the system tends to collapse back to the original state of the mixture.

On top of these considerations, additional comments can be made: First, multiple different free energy landscape topologies can be obtained depending on the thermodynamic parameter set. Accordingly, the AAPS-like region in Fig.\ref{fig:Landscape}-b can overlap in various manners with the energy barrier curve and modify the volume fraction ranges dominated by each pathway. For instance, two-step nucleation is favoured over a significant composition range for the immiscible blends in Sec.\ref{Sec:AAPSCryst} and Sec.\ref{Sec:DiffLimSpino}. In any case, the amplitude of the fluctuations has to be sufficient to overcome the energy barrier threshold at some point of the process. In particular, this generally limits one-step nucleation from lower solute volume fractions since the associated curve increases with lower $\phi$ (see Fig.\ref{fig:Landscape}-b). Second, as already mentioned, nucleation via a two-step process can be critically hindered at the lowest concentrations. As a matter of fact, since the proportions of the intermediate phases respect a lever rule, it can become very difficult to start from low $\phi_0$ and generate domains which are simultaneously larger than the critical nucleus size and have an adequately high solute content. Third, kinetic effects also play a crucial role. In general, the exact path through the free energy landscape is determined by the respective timescales on which the transport phenomena occur. In extreme cases where the evolution of the volume fraction is very slow as compared to the order parameter $\psi$, two-step nucleation can be significantly impaired (see Sec.\ref{Sec:DiffLim}). Then, the system regresses rapidly to the amorphous state ($\psi = 0$) before AAPS-like demixing produces phases with favorable compositions for nucleation of stable crystals.

\section{General Crystallization Behavior in Binary Mixtures}\label{Sec:CrystallizationBehavior}

\subsection{Regular Crystalline Morphology Formation \& Reference Transformation Kinetics}\label{Sec:RegularCryst}

Following these preliminary discussions, simulations of crystalline morphology formation in binary mixtures are now addressed. For this purpose, all computations carried out in the context of this work rely on a 2D square box of 512 nm$^{\text{2}}$ with a grid spacing of $\Delta x = \Delta y = 1$ nm as simulation domain. Periodic boundary conditions are applied at each edge to model crystallization within the bulk of the blend. In order to optimize time-stepping, the time scheme makes use of an adaptive procedure that allows $\Delta t$ to reach the largest possible value that still ensures numerical convergence \cite{ronsin_phase-field_2022}.

\begin{table}[!b]
    \centering
    \caption{List of thermodynamic and kinetic parameters used for the initial reference simulation of the PCBM-oDCB binary blend. The scaling parameter for the self-diffusion coefficients ($\Lambda_0$) is set equal to 1 in all simulations for which no other value is explicitly specified (see Sec.\ref{Sec:SensitivityStudy}, Sec.\ref{Sec:DiffLim} and Sec.\ref{Sec:DiffLimSpino})}
    \label{tab:SimulationParameter}
    \begin{tabular}{|c|c|c|}
         \hline
         Parameter & Value & Units \\
         \hline
         $\rho$ & 1600 & kg/m$^{\text{3}}$  \\
         $v_0$ & 1.131$\times$10$^{\text{-4}}$ & m$^{\text{3}}$/mol \\
         $N_1$ & 5.0298  & /\\
         $N_2$ & 1 & /\\
         $T$ & 333 & K \\
         $T_m$ & 558 & K \\
         $L$ & 20000 & J/kg \\
         $W$ & 40322.5806 & J/kg \\
         $\chi_{aa}$ & 0.7248 & /\\
         $\chi_{ca}$ & 1.0836 & /\\
         $\varepsilon^2$ & 1$\times$10$^{\text{-10}}$ & J/m\\
         $\kappa$ & 2$\times$10$^{\text{-10}}$ & J/m\\
         $\alpha$ & 8.1621$\times$10$^{\text{7}}$ & J/m$^{\text{3}}$ \\
         $M_0$ & 0.1 & s$^{\text{-1}}$\\
         $\Lambda_0$ & 1 & / \\
         $D_{1|{\phi_0\rightarrow 0}}^{\mathrm{(self)}}$ & 5$\Lambda_0\times$10$^{\text{-10}}$ & m$^{\text{2}}$/s \\
         $D_{1|{\phi_0\rightarrow 1}}^{\mathrm{(self)}}$ & $\Lambda_0\times$10$^{\text{-13}}$ & m$^{\text{2}}$/s \\
         $D_{2|{\phi_0\rightarrow 0}}^{\mathrm{(self)}}$ & 2$\Lambda_0\times$10$^{\text{-9}}$ & m$^{\text{2}}$/s \\
         $D_{2|{\phi_0\rightarrow 1}}^{\mathrm{(self)}}$ & $\Lambda_0\times$10$^{\text{-12}}$ & m$^{\text{2}}$/s \\
         \hline
    \end{tabular}
    
\end{table}

\begin{figure*}[!htb]
    \centering
    \includegraphics[scale=0.28]{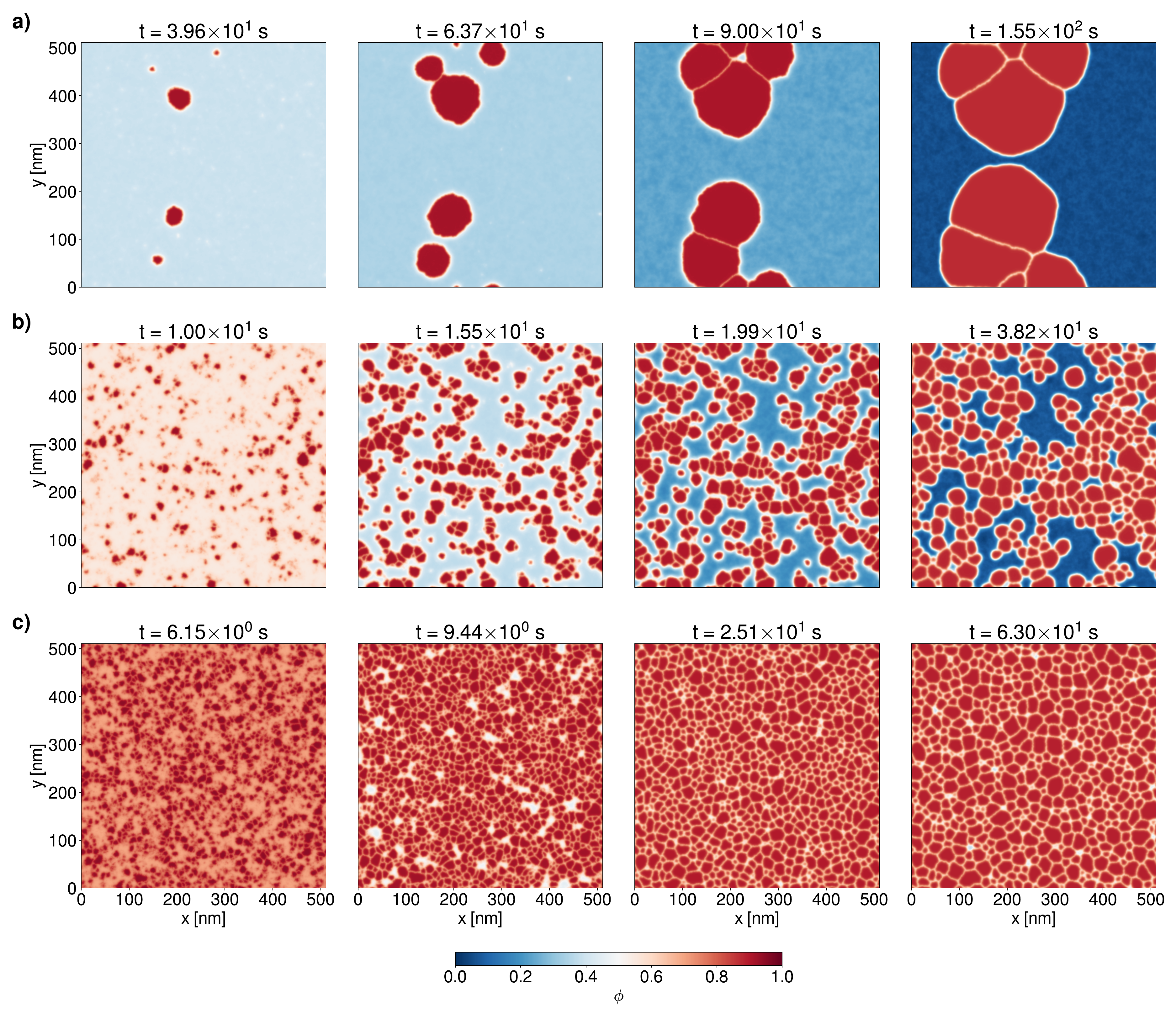}
    \caption{Progress of crystallization for the PCBM-oDCB binary system at $T$ = 333 K and for three different blend ratios (\textbf{a)} $\phi_0$ = 0.4, \textbf{b)} $\phi_0$ = 0.6, \textbf{c)} $\phi_0$ = 0.8). The volume fraction field $\phi$ of the crystallizing component is shown with a color code ranging from deep red (high $\phi$) to deep blue (low $\phi$). Thermodynamic and kinetic parameters used for the calculation are specified in Tab.\ref{tab:SimulationParameter}.}
    \label{fig:RegularCryst}
\end{figure*}

To begin the exploration of the different possible scenarios, the crystallization regime herein referred to as "regular" is considered, as it produces the most similar structures in comparison with crystallization from a pure material melt. In this configuration, the binary blend is miscible in the amorphous state and solute material is always available at the crystal growth front due to its relatively rapid diffusion towards interfaces. Separation of both chemical species only occurs once a crystal germ has formed. In addition, the inclusion rate of molecules into a crystal is assumed to be independent of the local mixture composition, i.e. $M(\phi) = M_0$. This constant mobility parameter $M_0$ should be calibrated with experimental crystallization kinetics data when specific material systems are to be replicated. Here, it is initially set to 0.1 $\text{s}^{\text{-1}}$, as this value (in combination with the other initial parameters) is suitable to realize the regular crystallization mode. The self-diffusion coefficients necessitated for calculation of $\Lambda(\phi)$ (see Eq.\ref{eq:Lambda} and Eq.\ref{eq:SelfD}) are taken over from earlier studies on the PCBM-oDCB solution \cite{ronsin_formation_2022}. In addition, the value of the impingement energy coefficient $\alpha$ is selected so that crystal interpenetration is prevented. The full set of thermodynamic, kinetic and numerical parameters used for the computations is summarized in Tab.\ref{tab:SimulationParameter}. 

\begin{figure*}[!htb]
    \centering
    \includegraphics[scale=0.54]{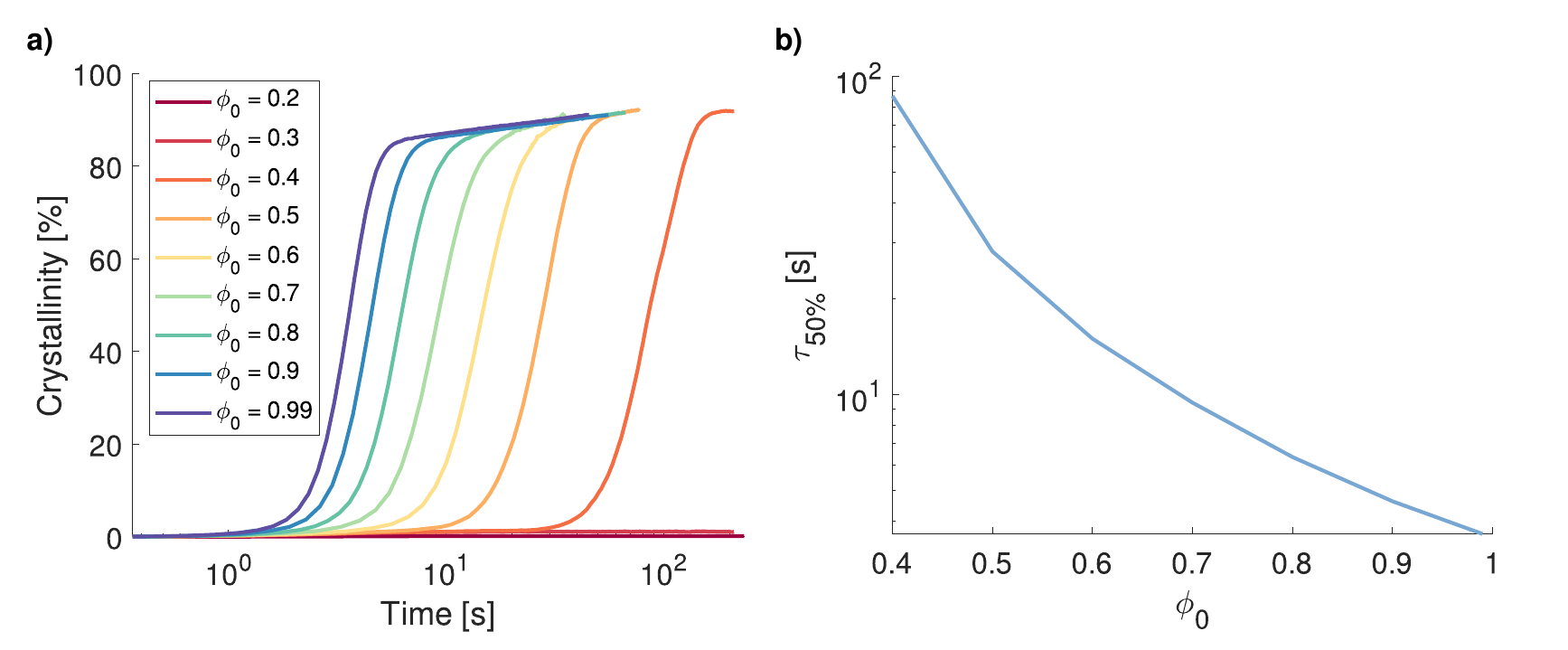}
    \caption{\textbf{a)} Transformation kinetics in the regular crystallization regime at different PCBM-oDCB blend ratios. \textbf{b)} Corresponding crystallization half-time $\tau_{50\%}$ as a function of blend composition.}
    \label{fig:CrystallizationKinetics}
\end{figure*}

Fig.\ref{fig:RegularCryst} compares the progress of crystallization for three different initial blend ratios ($\phi_0$ = 0.4, $\phi_0$ = 0.6, $\phi_0$ = 0.8). First, it can be observed that, for every simulated setup, three stages of crystallization, namely nucleation, free growth, and grain coarsening - i.e. the expansion of bigger crystals at the expense of smaller ones - are present. Apart from the final amount of crystallized material, which is straightforwardly related to the initial overall content of solute $\phi_0$, the simulations mainly differ in the extent in which these different mechanisms shape the crystallization process. More specifically, the blend ratio affects the balance between nucleation and growth. In this way, it can be remarked that morphology formation from lower $\phi_0$ tends to be more growth-dominated than from higher blend ratios - which, in turn, is nucleation-dominated. As can be visualized on the free energy surface contour plot (Fig.\ref{fig:Landscape}-b), the value of $\psi$ which corresponds to the location of the energy barrier increases with lower $\phi$. For this reason, less fluctuations are able to overcome the barrier. The associated critical nucleus size augments as well. As a consequence, stable nuclei are less likely to appear, leaving more solute material available for incorporation in already formed germs. In this configuration, the growth stage is favored compared to nucleation and only a limited number of relatively large crystals emerges. Conversely, with higher solute content, the energy barrier location drops to lower $\psi$. Accordingly, nucleation is facilitated up to the point where growth is hindered since most of the crystalline material is directly consumed for nuclei production. Crystals are then comparatively smaller (at least, before coarsening becomes significant).

Grain coarsening occurs predominantly at the end of the crystallization process, when most of the available solute material has effectively crystallized. It is driven by the minimization of surface tension contributions to the free energy of the system, which are most important at grain interfaces. This leads to a reduction of boundaries in between crystals. Remark that this coarsening mechanism is mathematically expressed by the gradient terms in Eq.\ref{eq:ACextended}. Another coarsening phenomenon that acts on phases of distinct compositions is captured by Eq.\ref{eq:CH} and will be more visible in an upcoming section (Sec.\ref{Sec:AAPSCryst}). From a general perspective, this shows that the proportions in which the species are initially mixed can lead to distinct formation regimes and eventually produce different structures with variations in crystal size distributions and spatial arrangements.

The crystallization kinetics can now further be characterized by recording the time-dependent evolution of the extent of transformed material relatively to the total amount of solute. This quantity is labeled as "crystallinity" in what follows and is compared for several initial blend ratios in Fig.\ref{fig:CrystallizationKinetics}-a. It can first be noticed that all curves present a sigmoidal shape, which is frequently observed for crystallization processes and usually modelled with the Johnson-Mehl-Avrami-Kolmogorov (JMAK) theory \cite{avrami_kinetics_1939} \cite{avrami_kinetics_1940} and its extensions \cite{kelly_modeling_2022}.

Moreover, it can be seen that the higher the global volume fraction of solute, the faster is the overall transformation. This is expected since the thermodynamic driving force for crystallization is proportional to $\phi L (T-T_m)/T$ (see Eq.\ref{eq:CrystallizationHomogeneous}). It is worth pointing out that, in this configuration, the coefficient for the local crystallization rate is kept constant ($M(\phi) = M_0$). For systems where diffusion properties vary significantly with the blend ratio, a related composition-dependence of $M(\phi)$ is anticipated to impact substantially the crystallization kinetics with different $\phi_0$. Further details about this are given in Sec.\ref{Sec:DilEn}.

Finally, recalling the discussion about nucleation pathways through the free energy landscape (Sec.\ref{Sec:Landscape}), it could be verified that no crystallization occurs within reasonable computation times at the lowest $\phi_0$. The reported volume fraction range thus starts at 0.2 in all forthcoming figures. Besides, even for higher mixing ratios that allow for direct one-step processes, the onset of crystallization can be considerably delayed because of the importance of the surface tension contributions (i.e. $\varepsilon^2$, and $W$) that impede nuclei formation. Thus, no crystallization is witnessed below $\phi_0$ = 0.4 within the timespan simulated in Fig.\ref{fig:CrystallizationKinetics}.

\subsection{Thermodynamic and Kinetic Sensitivity Study}\label{Sec:SensitivityStudy}

\begin{figure*}[!htb]
    \centering
    \includegraphics[scale=0.47]{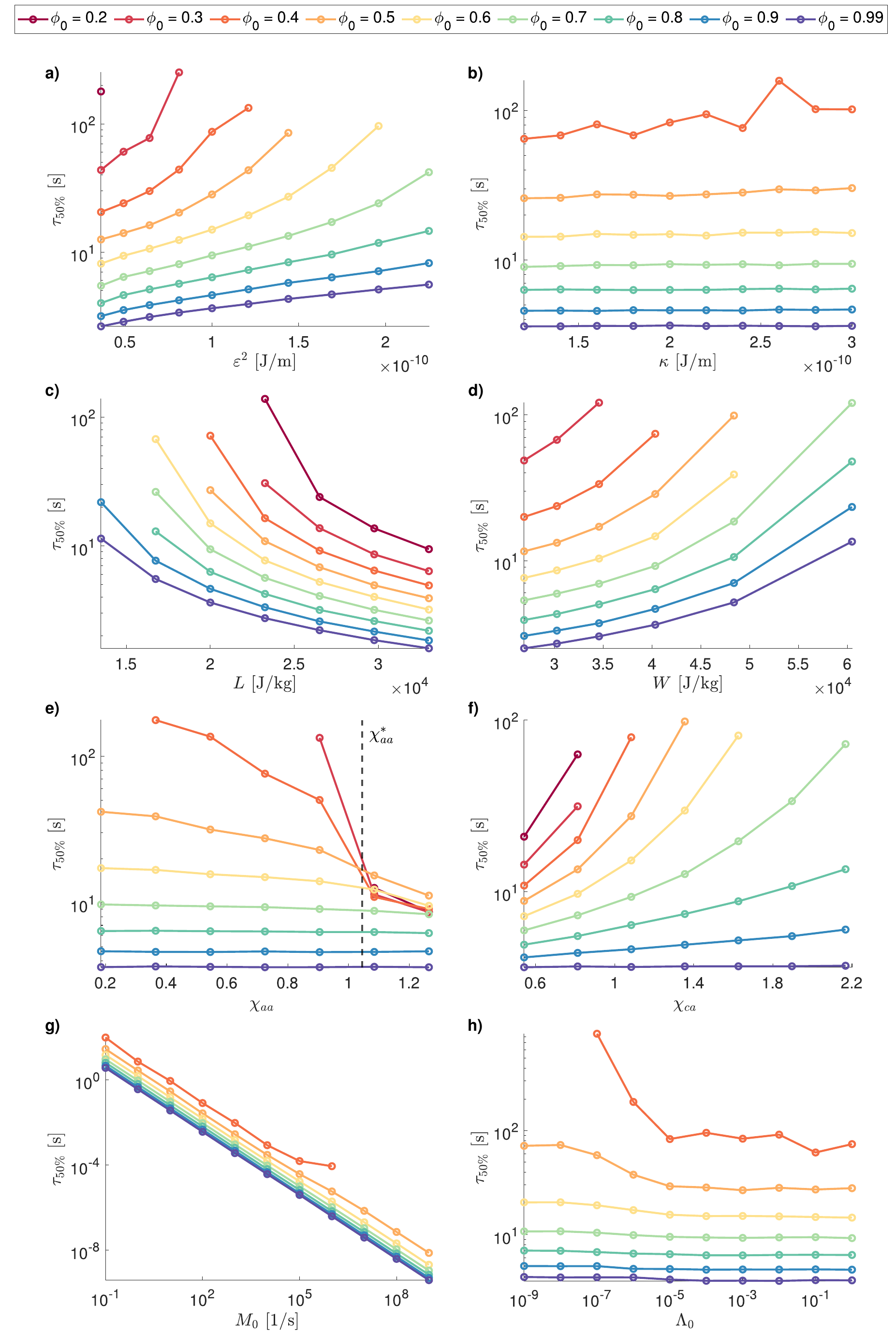}
    \caption{Sensitivity of the crystallization half-time $\tau_{50\%}$ to \textbf{a)-b)} surface tension, \textbf{c)-f)} thermodynamic and \textbf{g)-h)} kinetic model parameter variations for different initial blend ratios $\phi_0$. For each separate benchmark, all parameters were maintained constant at the values reported in Tab.\ref{tab:SimulationParameter}, except the variable specified on the abscissa.}
    \label{fig:SensitivityStudy}
\end{figure*}

Having analysed the crystallization behavior over the whole concentration range for a fixed binary system (Tab.\ref{tab:SimulationParameter}), the sensitivity of the process kinetics against thermodynamic and kinetic parameter variations is now investigated. To quantify this, the half-time of crystallization $\tau_{50\%}$ is defined as the time when the extent of crystallized material reaches 50\% of the total amount of available crystalline component. $\tau_{50\%}$ can be extracted from transformation plots such as Fig.\ref{fig:CrystallizationKinetics}-a. The overall crystallization rate can then be approximated by taking its inverse (i.e. 1/$\tau_{50\%}$). In order to carry out this study, all parameters from Tab.\ref{tab:SimulationParameter} are kept constant, except the indicated variable of interest, which is examined within the interval reported in Fig.\ref{fig:SensitivityStudy}. 

At first, the focus is restricted to surface tension parameters, namely $\varepsilon^2$ (Eq.\ref{eq:CrystallizationGradients}) and $\kappa$ (Eq.\ref{eq:MixingGradients}). As can be seen in Fig.\ref{fig:SensitivityStudy}-a and Fig.\ref{fig:SensitivityStudy}-b, $\varepsilon^2$ has globally a more severe influence on the crystallization kinetics as compared to $\kappa$. In addition, the general sensitivity is more important at low volume fractions for both coefficients. It can be remarked that the curves for the more growth-dominated low $\phi_0$ tend to fluctuate more due to the emergence of fewer nuclei, which results in poorer statistics for $\tau_{50\%}$. It is also noteworthy that the balance between both parameters affects which of the surface tensions arising at interfaces between a crystalline region and its amorphous surroundings, or at boundaries separating two distinct crystal grains, is the strongest. Accordingly, crystals tend to agglomerate or spread across the domain, depending on which associated energy contribution is the highest.

Next, the thermodynamic parameters $L$, $W$, $\chi_{ca}$ and $\chi_{aa}$ are varied. From Fig.\ref{fig:SensitivityStudy}-c it can be concluded that increasing the latent heat of fusion $L$ boosts crystallization. An increase of the energy barrier coefficient $W$ conversely inhibits it (Fig.\ref{fig:SensitivityStudy}-d). This is expected since the former parameter determines the thermodynamic driving force for crystallization while the latter dictates the height of the energy barrier to be overcome for nucleation. Considering the logarithmic axes of the figures, it can be seen that both have an impact on the crystallization half-time that is stronger than an exponential trend. Notice also that the dependencies displayed here for $L$ do not involve the coupling with $\chi_{ca}$ (Eq.\ref{eq:CAInteraction}), that is $\chi_{ca}$ is kept constant. Although the spacing between the curves for different mixture compositions are not identical, the general behavior is nonetheless similar when $\chi_{ca}$ is assumed to vary with $L$ according to Eq.\ref{eq:CAInteraction} (see SI-D). 

Concerning the evolution of $\tau_{50\%}$ as a function of the amorphous-crystalline interaction parameter, it can be observed in Fig.\ref{fig:SensitivityStudy}-f that the half-time of crystallization is delayed with ascending $\chi_{ca}$. This is because the energetical cost due to enthalpic interactions within crystals is larger. It can be recognized that the effect is more pronounced for lower $\phi_0$. In contrast, the crystallization rate increases with $\chi_{aa}$ (Fig.\ref{fig:SensitivityStudy}-e). This is coherent in terms of energy balance since strong interactions between both species in the amorphous phase promote crystallization. Similarly to the results for $\chi_{ca}$, the impact is more significant at lower solute content. Past a certain $\chi_{aa}$ threshold, crystallization is furthermore abruptly accelerated and becomes even faster than for some of the higher concentrated solutions. This behavior is symptomatic for an AAPS occurring ahead of the crystallization process and will be detailed in an upcoming section (Sec.\ref{Sec:AAPSCryst}). To predict the appearance of the amorphous miscibility gap, the following analytical formula can be used \cite{rubinstein_polymer_2003}: 

\begin{equation}
    \chi_{aa}^* = \frac{1}{2}\left( \frac{1}{\sqrt{N_1}} + \frac{1}{\sqrt{N_2}} \right)^2 ~,
    \label{eq:CriticalChi}
\end{equation}

where $\chi_{aa}^*$ represents the critical value of the amorphous-amorphous interaction parameter beyond which the system might become immiscible depending on the blend ratio $\phi_0$. $N_1$ and $N_2$ also play a role in this equation. In the present context, $N_2$ is always equal to 1. Therefore, increasing the relative solute size $N_1$ in the Flory-Huggins lattice reduces miscibility as well and can eventually trigger AAPS before crystal nucleation (see SI-D). 

In the end, the influence of the kinetic parameters that account for molecular ordering and diffusion rates, respectively $M_0$ and $\Lambda(\phi)$ (Eq.\ref{eq:ACextended} and Eq.\ref{eq:CH}), remains to be analysed. Since $\Lambda$ is a function of $\phi$ (Eq.\ref{eq:Lambda}), it is scaled homogeneously by multiplication with a non-dimensional coefficient $\Lambda_0$. On the one hand, it is visible in Fig.\ref{fig:SensitivityStudy}-g that the crystallization half-time is inversely proportional to $M_0$ until a certain value ($M_0 \sim 10^4$ s$^{-1}$) beyond which the curves slightly deviate from the trend, especially for lower solute content. On the other hand, $\tau_{50\%}$ is almost insensitive to variations in diffusion properties above $\Lambda_0 \sim 10^{-5}$ (Fig.\ref{fig:SensitivityStudy}-h). Below this point the crystallization kinetics experience a slowdown, which is also more pronounced at lower $\phi_0$. 

Considering that the crystallization kinetics are bound simultaneously by the maximum speed at which molecules can attach to the surface of growing crystals (determined by $M_0$) and the rate at which solute material can diffuse from amorphous regions to crystal interfaces (given by $\Lambda(\phi)$), the ratio of both parameters, $M_0/\Lambda(\phi)$, establishes the governing equation that drives the crystallization process. When $M_0/\Lambda(\phi)$ is low enough, diffusion in the amorphous phase occurs on timescales smaller by several orders of magnitude in comparison to crystal growth, so that there is always sufficient material available at the growth front. Then, the crystal interface mobility $M_0$ is the limiting kinetic factor that dictates the crystallization rate. Accordingly, a variation in $\Lambda_0$ has no effect on the crystallization half-time in this case. Once this condition is not fulfilled anymore, i.e. at larger $M_0/\Lambda(\phi)$, crystal growth becomes diffusion-limited. Morphology formation properties specific to this crystallization regime will be examined in Sec.\ref{Sec:DiffLim}. Note that in Fig.\ref{fig:SensitivityStudy}-g crystallization is not recorded for $\phi_0 < 0.4$ even when $M_0$ increases. This is due to the limiting criterion for the time-step size which is inversely proportional to $M_0$ \cite{ronsin_phase-field_2022}. Thus, the maximum times in reach with reasonable computational effort also decrease with increasing $M_0$, which renders it difficult to access the timescales where the first nuclei appear. The $\tau_{50\%}$ curves for these compositions are nonetheless expected to be parallel to the other ones (at least before the deviation from the inversely proportional trend).

Finally, it can be pointed out that the effects of all parameters related to the Cahn-Hilliard equation (Eq.\ref{eq:CH}), and thus to volume fraction changes (i.e. $\chi_{aa}$, $\chi_{ca}$ and $\kappa$), become very limited for high solute volume fractions $\phi_0$ since, in that case, the composition remains nearly homogeneous over the transformation process. Moreover, it may be mentioned that nucleation is, in general, more heavily dependent on thermodynamic parameters than crystal growth. As a result, most of the related $\tau_{50\%}$ sensitivities presented above reflect predominantly the impact of parameters on the nucleation rate. Thus, this means that, in addition to being generally favoured at low solute content, crystal growth is also more dominant when nucleation is penalized due to high $\varepsilon$, $W$, $\chi_{ca}$, or conversely low $L$ and $\chi_{aa}$ values.

\section{Review of Noteworthy Crystalline Morphology Formation Regimes}\label{Sec:CrystallizationModes}

\subsection{Diffusion-Limited Crystallization}\label{Sec:DiffLim}

\begin{figure*}[!htb]
    \centering
    \includegraphics[scale=0.28]{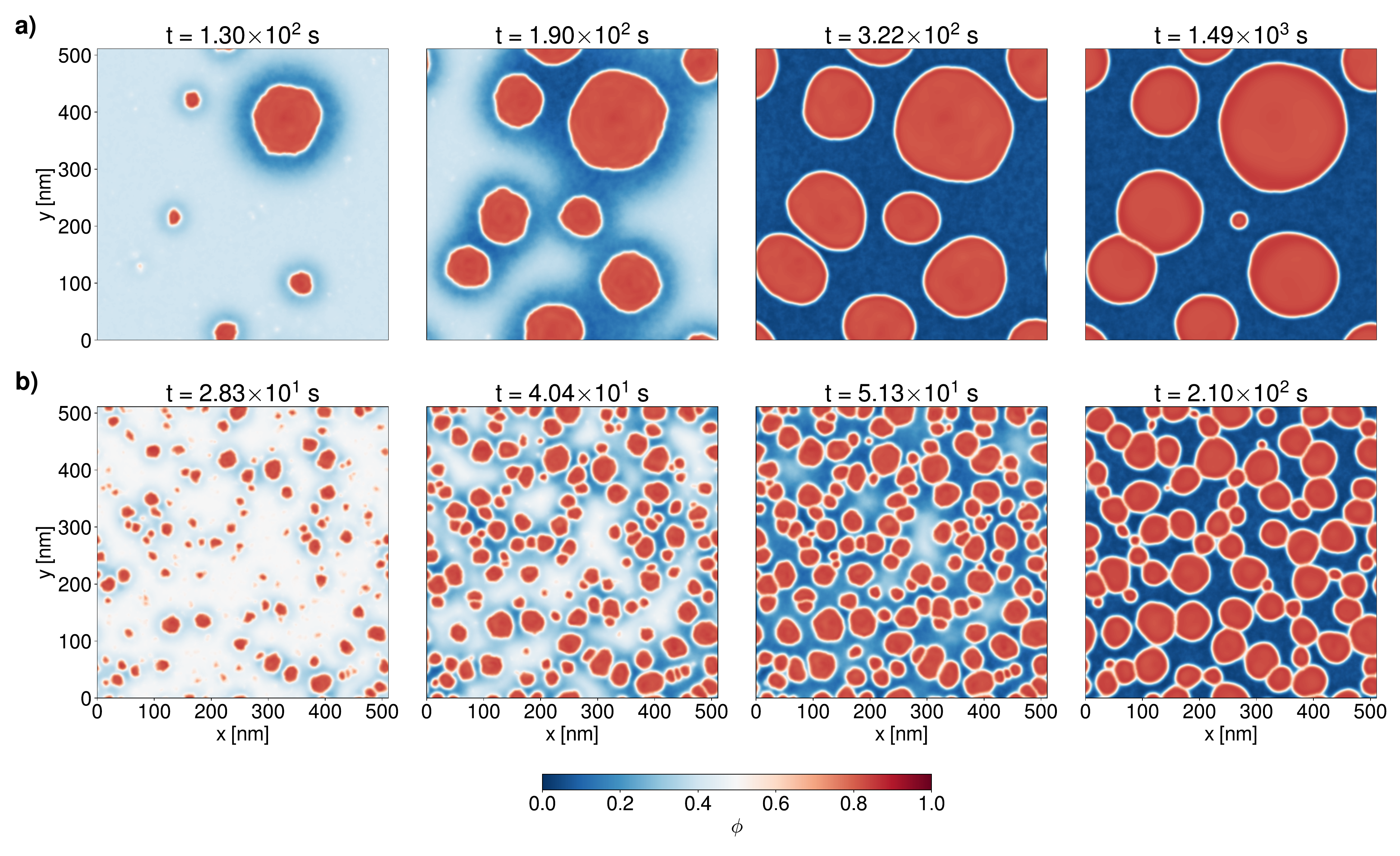}
    \caption{Progress of crystallization simulated for a diffusion-limited binary blend at $T$ = 333 K. The volume fraction field $\phi$ is shown at different times for \textbf{a)} $\phi_0$ = 0.4 and \textbf{b)} $\phi_0$ = 0.5. Diffusion rates within the blends are slowed down by a factor $\Lambda_0$ = 10$^{\text{-6}}$ in comparison to the reference simulation case (Fig.\ref{fig:RegularCryst}). Other thermodynamic and kinetic parameters are identical to those listed in Tab.\ref{tab:SimulationParameter}. The onset of the diffusion-limited regime is characterized by the appearance of depletion zones around growing crystals. At late crystallization stages, bead chain-like structures can also be observed.}
    \label{fig:DepletionZone}
\end{figure*}

In this section, the morphology formation is analysed for diffusion-limited systems. In the regular crystallization case, where $M(\phi)/\Lambda(\phi)$ is sufficiently low, diffusion within the amorphous phase is faster than crystal growth. As can be visualized in Fig.\ref{fig:RegularCryst} (Sec.\ref{Sec:RegularCryst}), this translates into a monotonously increasing volume fraction profile from the amorphous region to the crystalline one. In contrast, when molecular diffusion happens on larger timescales than crystal growth, i.e. comparatively higher $M(\phi)/\Lambda(\phi)$, solute material is missing at the growth front since crystals consume it faster than it can diffuse from the bulk of the amorphous phase to the interface. As is shown in Fig.\ref{fig:DepletionZone}, a depletion zone consequently emerges around growing crystals. The appearance of this feature indicates the onset of the diffusion-limited regime, sometimes also referred to as diffusion-controlled. The first repercussion of diffusion limitations on the transformation kinetics is the significant slowdown of the crystallization rate noticed in Fig.\ref{fig:SensitivityStudy}-h (Sec.\ref{Sec:SensitivityStudy} and SI-E). As mentioned before (Sec.\ref{Sec:RegularCryst}), systems with lower initial solute concentration are more growth dominated. Thus, and since diffusion-control affects mainly the growth stage, a decrease of the crystallization rate becomes more apparent towards low $\phi_0$. Additionally, the fact that crystalline material is inherently more scarce at low overall solute contents results in more pronounced concentration deficits at the growth front, which also contributes to this kinetic hindering of crystallization. 

\begin{figure*}[!htb]
    \centering
    \includegraphics[scale=0.28]{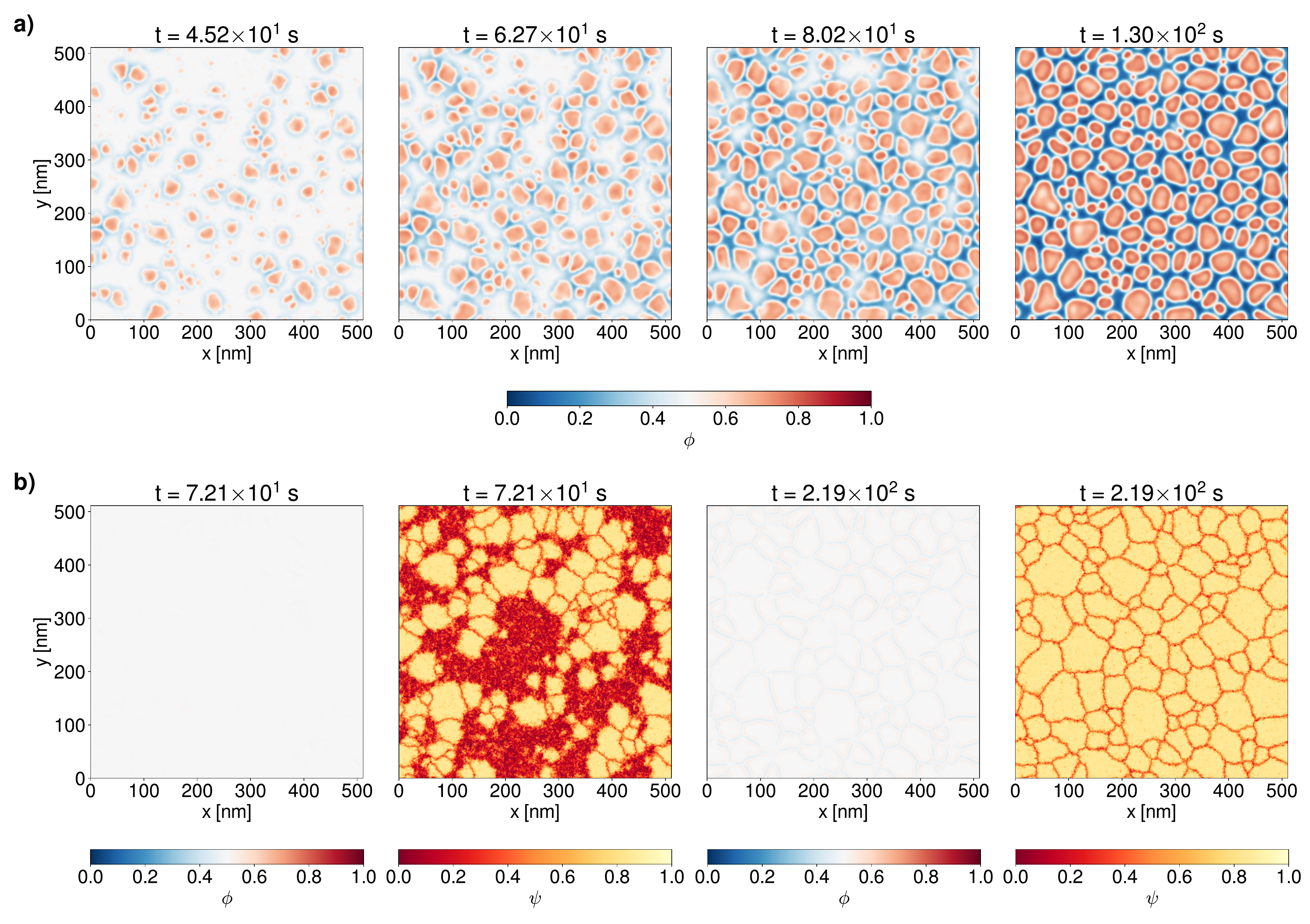}
    \caption{Progress of crystallization simulated for a diffusion-limited binary blend at $T$ = 333 K. The volume fraction ($\phi$) and order parameter ($\psi$) fields are shown at different times for $\phi_0$ = 0.5. Diffusion rates within the blend are slowed down by a factor \textbf{a)} $\Lambda_0$ = 10$^{\text{-7}}$ and \textbf{b)} $\Lambda_0$ = 10$^{\text{-9}}$ in comparison to the reference simulation case (Fig.\ref{fig:RegularCryst}). Other thermodynamic and kinetic parameters are identical to those listed in Tab.\ref{tab:SimulationParameter}. In \textbf{a)}, the relative strength of the diffusion limitations leads to heterogeneous composition profiles within forming crystals, which consequently adopt polygonal shapes. In \textbf{b)}, molecular inter-diffusion is substantially slower than solute ordering. As a result, impure crystals form with a composition that remains at the initial blend ratio $\phi_0$.}
    \label{fig:PolygonalCrystals}
\end{figure*}

Diffusion limitations impact not only the crystallization kinetics, but also the morphology formation within the blend. Crystals tend to become round in this regime: nucleation in the vicinity of a growing germ being disabled due to the absence of sufficient crystalline material, space is available to adopt the most energetically favorable shape. It can be noticed as well that the depletion zone eventually disappears after the termination of crystal growth, when the equilibrium compositions are reached homogeneously in both amorphous and crystalline phases. Moreover, if the surface tension properties are such that boundaries between adjacent crystals are energetically favoured over interfaces from ordered to amorphous phases, crystals are prone to stick together and bead chain-like structures can ultimately be observed in the late crystallization stages, as displayed in Fig.\ref{fig:DepletionZone}.

By increasing further the ratio $M(\phi)/\Lambda(\phi)$ the volume fraction profile progressively reacts with more delay to changes in order parameter $\psi$. In addition to depletion zones around growing germs, the volume fraction within crystallites becomes inhomogeneous (Fig.\ref{fig:PolygonalCrystals}-a). Indeed, it remains lower than the solidus equilibrium composition. Since a considerable amount of the second material is trapped within these impure crystals, the amount of amorphous phase is significantly lower. As a result, crystals also cover larger portions of the domain and adopt more polygonal forms. Upon coarsening, the grains become rounder but solute material inside crystals accumulates near their border rather than in the center. Although simulations were not pursued to larger timescales, it is expected that this conformation eventually breaks down and coarsens since the energetically optimal state is composed of one single crystalline droplet surrounded by the amorphous phase. The volume fractions of the phases at equilibrium should then be homogeneous, as predicted by the phase diagram (Fig.\ref{fig:PhaseDiagram}). 

Finally, the diffusion-limited regime with depletion zones cannot be sustained for the highest $M(\phi)/\Lambda(\phi)$. If the crystal interface mobility becomes too important relatively to solute diffusion, information about order parameter field fluctuations does not propagate through the volume fraction field on time, so that the composition homogeneously remains at its initial level. Two situations can be distinguished under these conditions. At lower overall content of crystalline material $\phi_0$, stable one-step nucleation is thermodynamically impossible without a change in composition (see Fig.\ref{fig:Landscape}-b in Sec.\ref{Sec:Landscape}). Two-step processes, along with common crystal purification, are kinetically hindered as well since AAPS-like demixing follows the diffusion kinetics. The system therefore stays in its original state without crystal formation for relatively long times. As a result, it can be seen in Fig.\ref{fig:SensitivityStudy}-g that no more crystallization is observed for low $\phi_0$ (here $\phi_0<0.4$) at the highest $M_0$ (or conversely at the lowest $\Lambda_0$ in Fig.\ref{fig:SensitivityStudy}-h). In contrast, at higher concentrations, one-step nucleation is permitted (however nearly without any composition alterations due to the aforementioned kinetic restrictions). $\psi$ values locally overcome the energy barrier and crystals form, but they remain highly impure (Fig.\ref{fig:PolygonalCrystals}-b). Equilibrium compositions and phase volume proportions can subsequently be achieved upon further grain boundary coarsening and crystal purification on considerably larger timescales.

\newpage

\begin{figure*}[!htb]
    \centering
    \includegraphics[scale=0.4]{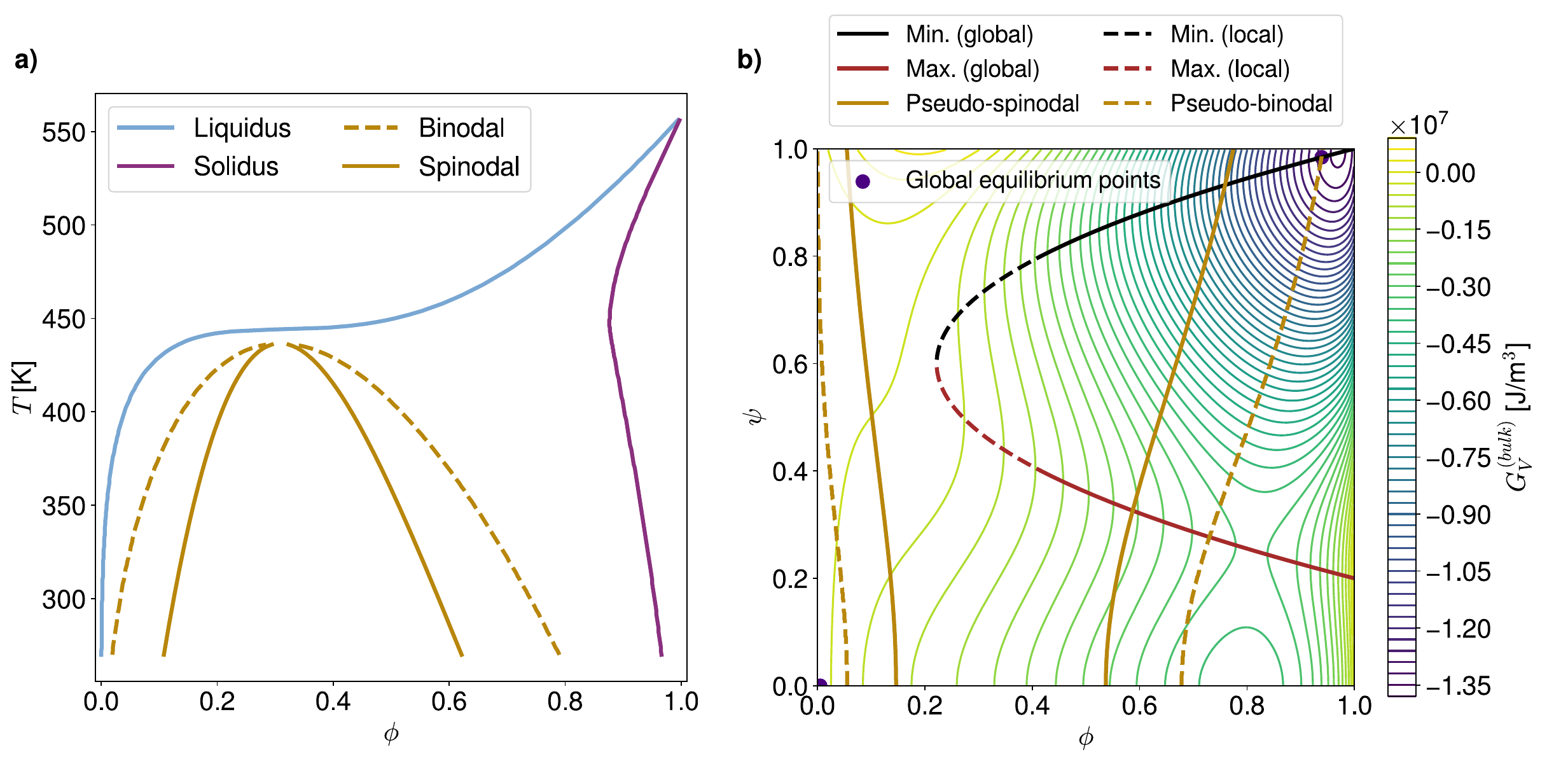}
    \caption{\textbf{a)} Phase diagram of an amorphous-crystalline system which exhibits an amorphous-amorphous immiscible UCST behavior below $T \simeq$ 440 K. The Flory-Huggins interaction parameter used for the calculation is given by $\chi_{aa}$ = 0.34 + 307.96/$T$. Other relevant thermodynamic parameters are identical to those referenced in Tab.\ref{tab:SimulationParameter}. \textbf{b)} Contour plot of the corresponding free energy landscape at $T$ = 333 K and $\chi_{aa}$ = 1.2648. }
    \label{fig:SpinodalPhaseDiagram}
\end{figure*}

\begin{figure*}[!htb]
    \centering
    \includegraphics[scale=0.28]{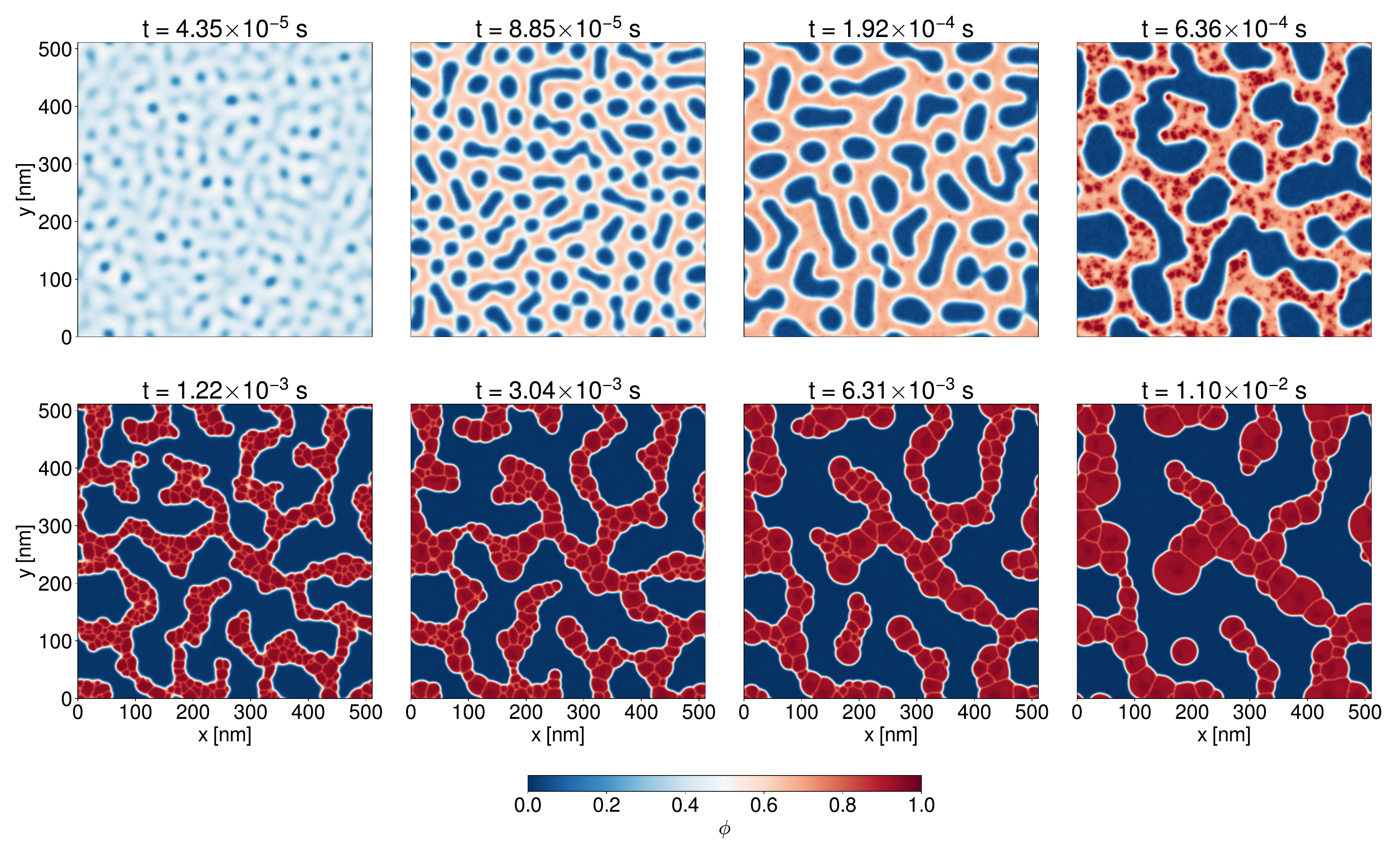}
    \caption{Progress of crystallization simulated for the immiscible system represented by the phase diagram Fig.\ref{fig:SpinodalPhaseDiagram}-a at $T$ = 333 K. The volume fraction field $\phi$ is shown at different times for $\phi_0$ = 0.4. The simulation parameters are identical to those listed in Tab.\ref{tab:SimulationParameter}, except $M_0$ = 1000 s$^{\text{-1}}$ and $\chi_{aa}$ = 1.2648. Following an initial spinodal decomposition, crystals nucleate and grow during the coarsening stage of the amorphous-amorphous phase separation (top row). The crystallization is favored in the domains where the solute is in majority, here the background matrix. Subsequently, grain coarsening also takes place after crystal impingement (bottom row).
    }
    \label{fig:Spino2}
\end{figure*}

\begin{figure*}[!b]
    \centering
    \includegraphics[scale=0.28]{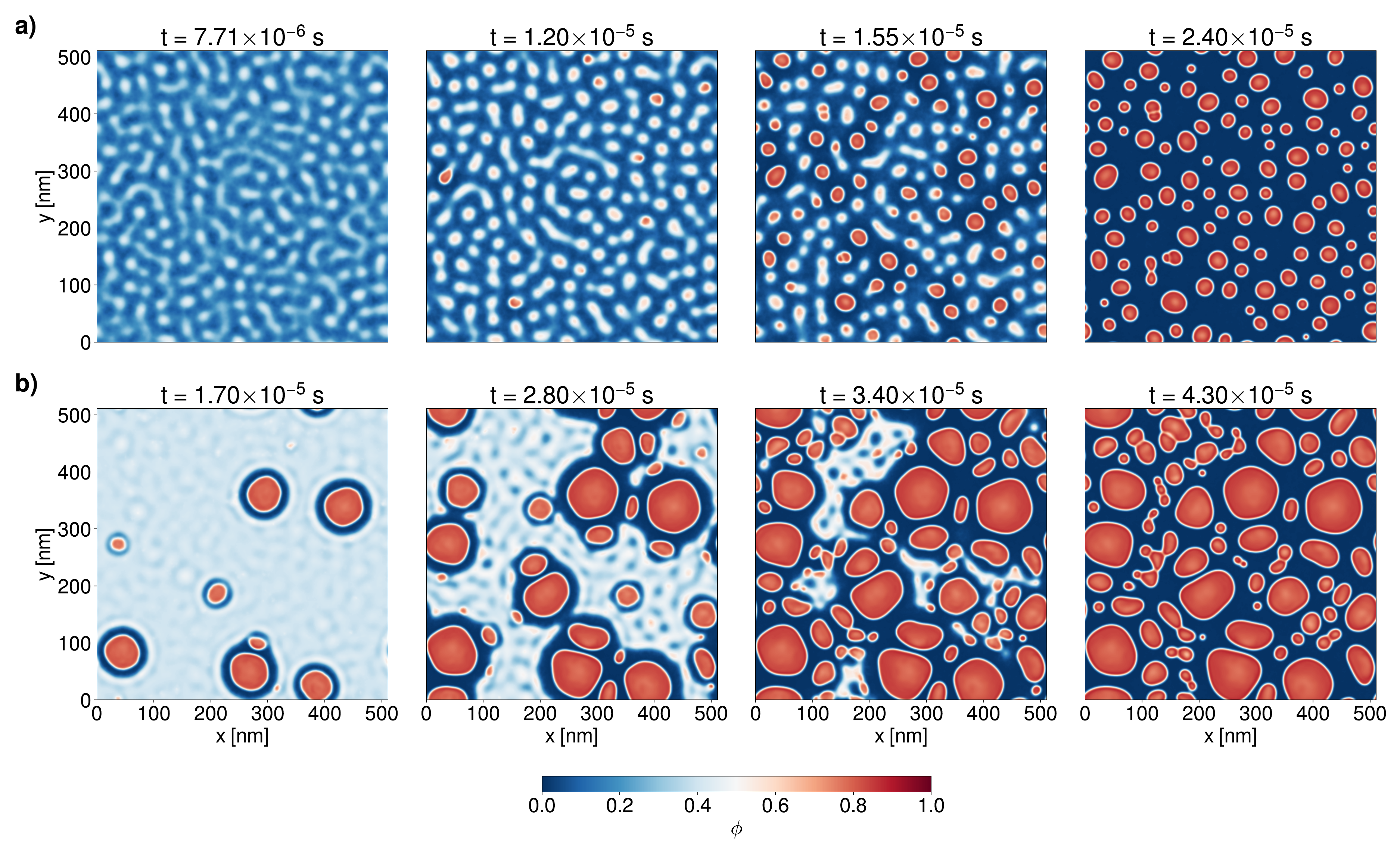}
    \caption{Progress of crystallization for an immiscible binary blend subject to diffusion limitations at $T$ = 333 K. The volume fraction field $\phi$ is shown at different times for blend ratios \textbf{a)} $\phi_0$ = 0.2 and \textbf{b)} $\phi_0$ = 0.4. For both simulations, the thermodynamic and kinetic parameters are identical to those listed in Tab.\ref{tab:SimulationParameter}, except $M_0$ = 10$^{\text{6}}$ s$^{\text{-1}}$ and $\chi_{aa}$ = 1.2648. In \textbf{a)}, a two-step crystallization process is undergone: the transformation is initiated by a spinodal decomposition followed by crystal nucleation and diffusion-controlled growth in the emerging solute majority phase. Conversely in \textbf{b)}, direct crystal nucleation and growth occurs while a parallel spinodal decomposition takes place in the remaining mixed amorphous phase. Due to diffusion limitations, the growing crystals are surrounded by depletion zones. The arising amorphous domains where solute is in majority present favorable conditions for secondary crystals to develop and the system progressively becomes more nucleation-dominated. Therefore, smaller and closer-spaced crystals form in the late stages. }
    \label{fig:Spino3}
\end{figure*}

\subsection{Demixing-Assisted Crystallization}\label{Sec:AAPSCryst}

Crystalline structures arising in immiscible mixtures that allow for a parallel AAPS process are studied in this section. All simulations shown here base on the thermodynamic parameters used to model the PCBM-oDCB system (Tab.\ref{tab:SimulationParameter}), by exception of $\chi_{aa}$ which is greater than $\chi_{aa}*$ = 1.0453 for the investigated temperature ($T$ = 333 K). The corresponding phase diagram and free energy landscape are reproduced in Fig.\ref{fig:SpinodalPhaseDiagram}-a and Fig.\ref{fig:SpinodalPhaseDiagram}-b, respectively. The unstable and metastable domains of the AAPS-like demixing region of the landscape intersect with the $\phi$-axis for $\phi$ $\in$ [0.146, 0.537] and $\phi \in$ [0.055, 0.678]. The endpoints of these intervals correspond to the classical amorphous-amorphous spinodal and binodal limits (at 333 K) which are also reported in the phase diagram (Fig.\ref{fig:SpinodalPhaseDiagram}-a), revealing a UCST amorphous miscibility gap. For blend ratios that lie within these ranges, accelerated nucleation can be triggered if the time required for AAPS \cite{konig_two-dimensional_2021} is shorter than the time for crystallization from the homogeneous mixture at $\phi_0$. In this case, amorphous demixing takes place first and generates domains with a solute content higher than $\phi_0$, as expected from the binodal line. As a consequence, crystallization preferably occurs in this solute majority phase at the rate - and with the nucleation/growth balance - that is associated with this latter composition. Due to the drastic reduction in crystallization time when AAPS develops beforehand crystallization (see Fig.\ref{fig:SensitivityStudy}-e in Sec.\ref{Sec:SensitivityStudy} and SI-F), this regime is qualified as "demixing-assisted" in what follows.

In this crystallization mode, substantially different types of morphologies develop as compared to the cases investigated in the precedent sections. Fig.\ref{fig:Spino2} depicts a typical formation pathway for this scenario. In the early stages of the process, AAPS (here SD) is triggered followed by coarsening. Hence, domains produced by the initial phase separation purify and tend to arrange as droplets of one component majority phase within a background matrix of the other. Volume proportions of these intermediate phases can be computed according to the lever rule applied to the amorphous binodal points of the phase diagram. In a second step, nuclei materialize in the domains of high solute content. The appearance of crystals during the coarsening partially quenches the AAPS pattern, so that it strongly determines the achieved crystalline structure. In the provided example, crystallization takes place in the background matrix, when wormlike elongated droplets of the solvent majority phase are still present. This results in the pseudo-bicontinuous conformation that can be seen at intermediate stages in Fig.\ref{fig:Spino2}. Note however that multiple other geometries are possible, depending on the blend ratio, the SD or NG nature of the demixing, the quench depth for AAPS (i.e. $\chi_{aa}-\chi_{aa}^*$ \cite{konig_two-dimensional_2021}) and the coarsening time before nucleation. Additional cases where the blend ratio and the nature of the demixing are varied can be visualized in the supplementary information (SI-G). Once crystals grow, further purification of the domains in terms of solvent and solute material begins since the final equilibrium compositions are given by the liquidus and solidus in the phase diagram (blue and purple lines on Fig.\ref{fig:SpinodalPhaseDiagram}-a). A change in phase volume proportions is also predicted by the lever rule between these latter curves. As a result, it can be noticed that the crystallized solute majority domains cover less space than the amorphous background matrix they are issued from. In contrast, the remaining solvent-rich amorphous region widens. Finally, on larger timescales, grain coarsening takes effect: surface tension contributions to the free energy are minimized, crystal interfaces suppressed, and thus percolating pathways tend to rupture.

\begin{figure*}[!b]
    \centering
    \includegraphics[scale=0.54]{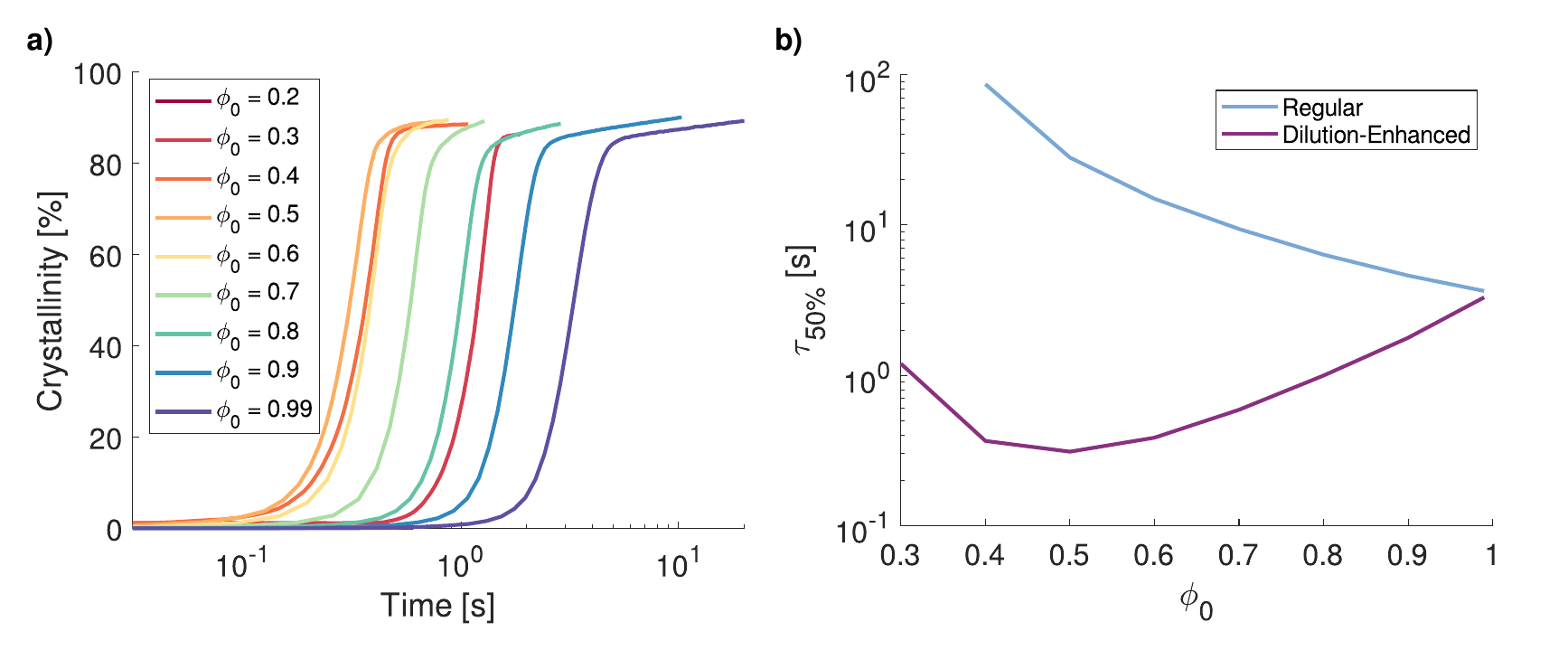}
    \caption{\textbf{a)} Transformation kinetics in the dilution-enhanced crystallization regime at different solvent-solute blend ratios. \textbf{b)} Comparison of the crystallization half-time as a function of blend composition in the dilution-enhanced and regular crystallization regimes.}
    \label{fig:DilutionEnhancedkinetics}
\end{figure*}

\subsection{Diffusion-Limited Crystallization in Immiscible Blends}\label{Sec:DiffLimSpino}

Special types of morphologies can also develop in immiscible systems where crystallization is fast in comparison to diffusion (i.e. $\chi_{aa}>\chi_{aa}*$ simultaneously with relatively large $M(\phi)/\Lambda(\phi)$). In that case, since AAPS kinetics are directly related to the diffusion rate, crystal nucleation and growth might proceed faster than amorphous-amorphous demixing. Different formation pathways can be distinguished under these circumstances. On the one hand, at low solute volume fractions (Fig.\ref{fig:Spino3}-a), where crystallization is only possible following a two-step nucleation route (see free energy landscape: Fig.\ref{fig:SpinodalPhaseDiagram}-b), AAPS still has to take place first. Crystals then start forming as soon as the solute content in the solute majority phase is high enough to allow for direct nucleation, even though the binodal compositions are not reached. Because the crystals grow surrounded with depletion zones due to diffusion limitations, wormlike structures that might have arisen during SD break up and the morphology evolves towards isolated, droplet-like crystals, which are ultimately subject to Ostwald ripening. 

On the other hand, if $\phi_0$ is large enough for a one-step process to happen, crystal nucleation and growth occurs before any amorphous phase dissociation (Fig.\ref{fig:Spino3}-b). Composition gradients within depletion zones trigger AAPS in a particular concentric pattern that is centered on crystals, as already observed in previous PF studies of placed initial crystal seeds in non-diffusion-controlled situations \cite{ronsin_role_2020}. Here, this effect is however confined to the vicinity of the crystals. The emerging high solute concentration rings that surround the depleted areas provide a preferential location for secondary germs to nucleate. Since depletion zones form as well around these new crystals, the rings with high solute content from which they originate are disrupted. Therefore, the secondary crystals still grow relatively isolated. In later stages, upon further progress of AAPS in the amorphous region, crystallization becomes more nucleation dominated in the solute majority phase. Thus, closer and smaller-sized germs materialize with a comparatively accelerated formation rate. Finally, after the solute material is fully consumed, the global morphology exhibits a relatively wide distribution in crystalline droplet sizes. In this way, those which appeared first under more growth-dominated conditions had time to become larger, adopt rounder shapes and are well-dispersed. In contrast, crystals which emerged last are smaller, closer to each other and slightly more distorted.

\subsection{Dilution-Enhanced Crystallization}\label{Sec:DilEn}

For all simulations presented in the previous sections, the local incorporation rate of amorphous solute into crystals is assumed constant over the whole composition range (i.e. $M(\phi) = M_0$). With the aim to model organic solvent-solute systems, it has nevertheless to be considered that the molecular mobility of crystalline materials such as polymers should be globally enhanced in the diluted state \cite{kim_modeling_2009}. Correspondingly, a faster crystal build-up should be kinetically enabled at lower solute volume fractions. To account for this, the value of the Allen-Cahn mobility parameter $M(\phi)$, which locally controls the amorphous-to-crystalline conversion rate, can be indexed as follows on the self-diffusion coefficient of the solute $D_1^{\mathrm{(self)}}$, which, in turn, depends on $\phi$ (see Eq.\ref{eq:SelfD} and SI-A) \cite{kim_modeling_2009} \cite{ronsin_phase-field_2022}:

\begin{equation}
    M(\phi) = \frac{D_1^{\mathrm{(self)}}(\phi)}{D_{1 |\phi_0 \rightarrow 1}^{\mathrm{(self)}}}M_0 ~.
    \label{eq:MDDep}
\end{equation}

The transformation kinetics observed in simulations implementing this modification are depicted in Fig.\ref{fig:DilutionEnhancedkinetics}. In comparison with the regular case (Fig.\ref{fig:CrystallizationKinetics}), the process does not monotonously accelerate with higher solute volume fractions, but rather exhibits a maximal crystallization rate at intermediate $\phi_0$. This is explained by the presence of two driving mechanisms for crystallization, which possess opposite trends with respect to solute content. In this way, the thermodynamic driving force (Eq.\ref{eq:CrystallizationHomogeneous}), that becomes stronger for high solute concentrations, competes with the molecular mobility enhancement that is more effective for high dilutions (i.e. low $\phi$) (Eq.\ref{eq:MDDep}). The exact solute volume fraction for which the crystallization rate is maximized then depends on material specific properties such as the heat of fusion or the self-diffusivities. 

Furthermore, morphological features arising in this crystallization scenario are qualitatively similar to those shown in Fig.\ref{fig:RegularCryst} for the regular structure formation pathway, even though the crystal build-up is less homogeneous due to local variations of $M$ with $\phi$. The nucleation rate at lower solute concentrations is also increased due to the higher values of $M(\phi)$. If the system is simultaneously subject to composition-dependent $M(\phi)$ (Eq.\ref{eq:MDDep}) and diffusion-limitations, or immiscibility in the amorphous state, depletion zones and crystalline structures shaped by an AAPS are also obtained. A summary of the corresponding morphologies is provided in the supplemental information (SI-H). Note that when the kinetic enhancement of the crystallization rate with dilution (Eq.\ref{eq:MDDep}) couples with these other physics, the overall process kinetics become even more sensitive to the material properties and substantial deviations from the typical transformation plots of Fig.\ref{fig:DilutionEnhancedkinetics} are expected.

\section{Discussion - Comparison with Experimental Results for OPV Systems}\label{Sec:Discussion}

In order to evaluate how the present PF approach can be applied to solution-processed PALs, reports discussing the phenomenology of crystallization in material blends relevant for OPV are drawn into comparison. It is found that the simulation outcomes are in line with many experimental results. To begin with, amorphous-crystalline polymer-polymer mixtures are known to exhibit transformation kinetics similar to the sigmoids presented in Fig.\ref{fig:CrystallizationKinetics} where the crystallization rate increases with the overall content of crystalline material \cite{mandelkern_crystallization_2004}. It is worth mentioning that, for these particular solute-solute configurations, diffusion properties should not vary significantly with the blend ratio. 

In addition, two-step nucleation processes where crystals appear within previously formed amorphous solute-rich clusters have been theorized and observed in protein, as well as in other organic, inorganic, and colloidal solutions \cite{erdemir_nucleation_2009} \cite{gvekilov_two-step_2010}, which agrees with predictions from the current PF model. The accelerated crystallization kinetics caused by an initial amorphous demixing are also consistent with publications of experiments on polymer-polymer blends \cite{tsuburaya_crystallization_2004} \cite{sugeno_ucst_2020} \cite{takamatsu_cooperative_2022}. It has been suggested that the rate of transformation is enhanced due to so-called up-hill diffusion of polymer chains into their corresponding majority phase where the thermodynamic driving force for crystallization is stronger. This corresponds to the mechanism observed in the present simulations. 

Note again that for these systems the diffusion coefficients are expected to be comparable throughout the whole volume fraction range. Evidence for dilution-enhanced crystal growth rates has also been reported for polymer solutions \cite{mandelkern_polymerdiluent_2004}. However, it is specified that the dependency of the crystallization kinetics on composition can vary substantially with the investigated material system, as the change with $\phi$ of other blend properties (e.g. miscibility, interdiffusivity, glass transition temperature, crystal growth dimensionality, etc.) is strongly determined by the selected materials. Thereby, no general composition-crystallization rate relationship could be derived. This is in agreement with the fact that the progress of crystallization can be significantly altered when molecular mobility enhancements due to dilution are combined with diffusion-control or AAPS physics, for instance, as it has been witnessed in this work.

Regarding particular applications for OPV, mixtures involving isotropically crystallizing species, such as PCBM in the PCE11-PCBM system that was visualized by Levitsky et al. using optical micrographs and vapor phase infiltration enhanced scanning electron microscopy \cite{levitsky_bridging_2021}, or the MDMO-PPV:PCBM blends imaged by Hoppe and co-workers with atomic force microscopy and scanning electron microscopy \cite{hoppe_nanoscale_2004} \cite{hoppe_nano-crystalline_2005} \cite{hoppe_morphology_2006} \cite{troshin_70fullerene-based_2011}, feature morphological structures that are similar to those reproduced in this work and provide encouraging perspectives for further investigations with the PF model as it is presented in this study. As mentioned before, anisotropic crystal growth, which is expected for a vast class of organic semi-conducting materials \cite{wang_organic_2018}, can be accounted for as well with this type of modelling framework \cite{granasy_phase-field_2014}.

Furthermore, the existence of depleted areas around crystals was evidenced with energy-filtered transmission electron microscopy for solvent-vapor annealed DRCN5T:PCBM all small molecule PALs by Harrei{\ss} et al. \cite{harreis_understanding_nodate}. Analyses with the current PF method can typically help in this case to determine whether diffusion-limited crystallization is responsible for the formation of such structures. Diffusion properties in the P3HT-PCBM model system leading to depletion zones around PCBM crystals were also studied by Berriman et al. \cite{berriman_molecular_2015}. Moreover, Yu et al. \cite{yu_diffusion-limited_2019} commented on the relevance of crystallization in strongly diffusion-controlled configurations for OPV systems. The importance of the changes in crystalline morphology encountered as the crystal growth rate becomes faster than the molecular diffusion rate has additionally been discussed by Wang et al. \cite{wang_coupling_2021} alongside with corresponding post-treatment strategies to adopt in order to optimize the resulting PALs for PCE. Insights from simulations can be of use here to identify when diffusion limitations become detrimental (or beneficial) for device performance, so that appropriate processing measures are undertaken.

Finally, it is recognized that the interplay of AAPS and crystallization is crucial for BHJ manufacturing \cite{wang_coupling_2021}. In general, OPV systems are likely to enter the miscibility gap of the phase diagram as composition and temperature significantly change during processing \cite{mcdowell_solvent_2018} \cite{peng_materials_2023}. Thus, several studies pointed out the significance of tailoring the components' miscibility for amorphous-crystalline morphology optimization \cite{ye_miscibilityfunction_2018}, the effects of spinodal decomposition on subsequent crystallization during thermal annealing \cite{levitsky_bridging_2021}, or issues with non-fullerene acceptor phase separation and pre-aggregation in solution \cite{xue_kinetic_2022} \cite{yao_preaggregation_2022}. As demonstrated in Sec.\ref{Sec:AAPSCryst}, complementary numerical investigations can be carried out with the present approach to gain a more detailed understanding of the competition between the physical phenomena involved in these processes.

\section{Conclusions}\label{Sec:Conclusion}
In conclusion, the crystallization behavior in binary, non-evaporating, amorphous-crystalline mixtures was assessed with the help of numerical simulations. More specifically, the objective was to gain a detailed comprehension of possible crystalline structure formation mechanisms based on the governing physics captured by the presented Phase-Field framework for solution-processed photoactive layer simulation. In this way, several mixture properties such as the blend ratio, the strength of thermodynamic driving force for crystallization, the height of the nucleation barrier, the surface tension, the miscibility of the mixed species, or their inter- and self-diffusivities, were found to play a crucial role in the investigated morphology formation process. Upon exploration of the thermodynamic and kinetic parameter space, the sensitivity of the transformation kinetics was evaluated and various crystallization scenarios were achieved. A more exhaustive analysis of the realized diffusion-limited, demixing-assisted and dilution-enhanced crystallization regimes revealed distinct morphology formation pathways with specific process kinetics, crystal size distributions, spatial arrangements and nucleation/growth balances. Thereby, remarkable morphological features such as depletion zones forming around crystals or percolating, bicontinous domains were produced as well. Moreover, the interplay of physical phenomena, which would separately lead to different crystallization modes, was also observed to result in unique geometrical configurations. Finally, experimental results reported in the literature were drawn into comparison and found to be in good qualitative agreement with numerical simulation outcomes.

Thus, the employed phase-field modelling approach demonstrates highly promising perspectives for understanding the interaction between complex nanoscale processes that take place simultaneously during photoactive film fabrication. It is expected that theoretical insights provided by simulations will substantially help the identification of process-structure relationships and support the derivation of guidelines for optimal manufacturing conditions. In order to increase even further the prediction range for morphological features that might arise during the solution-deposition process, supplementary physical phenomena are anticipated to be of importance. Future code developments therefore include extensions for anisotropy, semi-crystallinity, substrate interactions and chemical reactions. In addition, upcoming investigations will concentrate on the validation of model predictions against experimental measurements for specific state-of-the-art material systems. To this end, particular attention will be dedicated to accurate input parameter acquisition and in-situ morphological property inspection with adequate thin-film characterization techniques. This will ultimately allow for reliable simulations of the complete photoactive film formation upon drying, which involves concurrent crystallization, demixing, evaporation and hydrodynamic processes.

\section*{Author Contributions}

M. Siber: Methodology, Investigation, Formal Analysis, Data Curation, Writing - Original Draft, Writing - Review and Editing.\\O. J. J. Ronsin: Conceptualization, Software, Supervision, Writing - Review and Editing.\\J. Harting: Project Administration, Funding Acquisition, Writing - Review and Editing. 

\section*{Conflicts of interest}
There are no conflicts to declare.


\section*{Acknowledgements}
The authors acknowledge financial support by the German Research Foundation (DFG, Project HA 4382/14-1), the European Commission (H2020 Program, Project 101008701/EMERGE), and the Helmholtz Association (SolarTAP Innovation Platform).

\section*{Data Availability}
In compliance with the regulations for projects funded by the German Research Foundation (DFG), the simulation data used for this article is made publicly accessible (see DOI 00.0000/00000000).



\balance


\bibliography{Morphology_Formation_in_Binary_Blends} 
\bibliographystyle{rsc} 


\end{document}